\documentclass[twoside,draft]{article}

\usepackage{amsfonts}
\usepackage{stmaryrd}
\usepackage{amssymb}
\usepackage{euscript}
\usepackage{amsthm}
\usepackage{amsmath}
\usepackage{amscd}
\usepackage{latexsym}
\usepackage{mathrsfs}
\usepackage{graphicx}
\usepackage{color}
\usepackage{dsfont}
\usepackage{bm}
\usepackage{cleveref}
\usepackage{enumitem}%

\numberwithin{equation}{section}

\newtheorem{theorem}{Theorem}[section]
\newtheorem{lem}{Lemma}[section]
\newtheorem{pro}{Proposition}[section]
\newtheorem{cor}{Corollary}[section]
\newtheorem{rem}{Remark}[section]
\newtheorem{rems}{Remarks}[section]
\newtheorem{ex}{Example}[section]
\newtheorem{defi}{Definition}[section]
\newtheorem{hyp}{Assumption}[section]

\newcommand{\bt}{\begin{theorem}}
\newcommand{\et}{\end{theorem}}
\newcommand{\bl}{\begin{lem}}
\newcommand{\el}{\end{lem}}
\newcommand{\bp}{\begin{pro}}
\newcommand{\ep}{\end{pro}}
\newcommand{\bcor}{\begin{cor}}
\newcommand{\ecor}{\end{cor}}

\newcommand{\bd}{\begin{defi} \rm }
\newcommand{\ed}{\end{defi}}
\newcommand{\brem }{\begin{rem} \rm }
\newcommand{\erem }{\end{rem}}
\newcommand{\brems }{\begin{rems} \rm }
\newcommand{\erems }{\end{rems}}
\newcommand{\bhyp }{\begin{hyp} \rm }
\newcommand{\ehyp }{\end{hyp}}
\newcommand{\bex}{\begin{ex} \rm }
\newcommand{\eex}{\end{ex}}

\newcommand{\yxi}{\xi}
\newcommand{\bxi}{\bar{\xi}}

\newcommand{\seq}[1]{{\lbrace #1 \rbrace}}

\newcommand{\bigcdot}{\mathbin{{\hbox{\scalebox{.65}{$\bullet$}}}}}
\newcommand{\bcd}{\bigcdot}

\newcommand{\wt}{\widetilde}
\newcommand{\wh}{\widehat}
\newcommand{\ov}{\overline}

\newcommand{\I}{\mathds{1}}

\newcommand{\llb}{\llbracket}
\newcommand{\rrb}{\rrbracket}

\newcommand{\esssup}{\operatornamewithlimits{ess\,sup}}
\newcommand{\essinf}{\operatornamewithlimits{ess\,inf}}

\newcommand{\vt}{\vartheta}
\newcommand{\whsigma}{\widehat{\sigma}}

\newcommand{\STF}{{\cT}(\FF)}
\newcommand{\STFt}{{\cT}_{t,T}(\FF)}
\newcommand{\STFp}{{\cT}^p(\FF)}

\newcommand{\bSTF}{\overline{\cT}(\FF)}
\newcommand{\bSTFt}{\overline{\cT}_{t,T}(\FF)}
\newcommand{\bSTFp}{\overline{\cT}^p(\FF)}

\newcommand{\STG}{{\cT}(\GG)}
\newcommand{\STGt}{{\cT}_{t,T}(\GG)}

\newcommand{\xx}{x}
\newcommand{\VV}{V}
\newcommand{\xS}{S}
\newcommand{\bcK}{\ov{\cK}}

\newcommand{\barS}{\ov{S}}

\newcommand{\barV}{\ov{V}}
\newcommand{\barY}{\ov{Y}}
\newcommand{\barZ}{\ov{Z}}

\newcommand{\whMM}{M^{\QQ}}
\newcommand{\QQphi}{\QQ^{\varphi}}
\newcommand{\EQphi}{\mathbb{E}_{\mathbb{Q}^{\varphi}}}

\newcommand{\whVu}{\wh{V}^u}
\newcommand{\VVu}{V^u}
\newcommand{\vv}{v}

\newcommand{\whX}{\wh{X}}
\newcommand{\whY}{\wh{Y}}
\newcommand{\whV}{\wh{V}}

\newcommand{\whtau}{\widehat{\tau}}

\newcommand{\wtY}{\wt{Y}}
\newcommand{\wtZ}{\wt{Z}}
\newcommand{\wtK}{\wt{K}}

\newcommand{\wtG}{\wt{G}}
\newcommand{\wtH}{\wt{H}}

\newcommand{\mG}{m^{\GG}}
\newcommand{\nG}{n^{\GG}}
\newcommand{\wtGamma}{\widetilde{\Gamma}}

\newcommand{\wtGphi}{\wt{G}^{\varphi}}

\newcommand{\wtGaphi}{\wt{\Gamma}^{\varphi}}

\newcommand{\omG}{m^{\GG,\QQphi}}

\newcommand{\oLambda}{\Lambda}
\newcommand{\oGamma}{\widetilde \Gamma^{\QQphi}}

\newcommand{\sTr}[1]{\mathscr{T}(#1)}
\newcommand{\cTr}[1]{\mathcal{T}(#1)}

\newcommand{\Mxt}{M^{x(\tau)}}

\newcommand{\cH}{\mathcal{H}}
\newcommand{\cS}{\mathcal{S}}
\newcommand{\cK}{\mathcal{K}}

\newcommand{\cOff}{\mathcal{O}(\FF)}
\newcommand{\cOgg}{\mathcal{O}(\GG)}

\newcommand{\cPff}{\mathcal{P}(\FF)}
\newcommand{\cPgg}{\mathcal{P}(\GG)}

\newcommand{\cOffd}{\mathcal{O}_d(\FF)}

\newcommand{\cPffd}{\mathcal{P}_d(\FF)}

\newcommand{\cPrff}{\mathcal{P}r(\FF)}

\newcommand{\cPrffd}{\mathcal{P}r\!_d(\FF)}

\newcommand{\Martf}{\mathcal{M}(\FF)}
\newcommand{\Martg}{\mathcal{M}(\GG)}

\newcommand{\Martft}{\mathcal{M}^{\vt}(\FF)}

\newcommand{\Martfl}{\mathcal{M}_{loc}(\FF)}

\newcommand{\Martflt}{\mathcal{M}^{\vt}_{loc}(\FF)}
\newcommand{\Martglt}{\mathcal{M}^{\vt}_{loc}(\GG)}

\newcommand{\cA}{{\mathcal A}}
\newcommand{\cE}{{\mathcal E}}
\newcommand{\cF}{{\mathcal F}}
\newcommand{\cG}{{\mathcal G}}

\newcommand{\cT}{{\mathcal T}}

\newcommand{\cQ}{{\mathcal Q}}
\newcommand{\cZ}{{\mathcal Z}}

\newcommand{\FF}{{\mathbb F}}
\newcommand{\GG}{{\mathbb G}}
\newcommand{\HH}{{\mathbb H}}
\newcommand{\RR}{{\mathbb R}}
\newcommand{\QQ}{{\mathbb Q}}
\newcommand{\PP}{{\mathbb P}}
\newcommand{\NN}{{\mathbb N}}
\newcommand{\EE}{{\mathbb E}}
\newcommand{\EP}{\mathbb{E}_{\mathbb{P}}}
\newcommand{\EQ}{\mathbb{E}_{\mathbb{Q}}}

\title{{\Large \bf VULNERABLE EUROPEAN AND AMERICAN OPTIONS IN A MARKET MODEL WITH OPTIONAL HAZARD PROCESS} \vskip 35 pt }

\author{Libo Li$\,^{a}$, Ruyi Liu$\,^{b}$ and Marek Rutkowski$\,^{b,c}$ \\ \\ \\ \\
\\ $^{a\,}$School of Mathematics and Statistics, University of New South Wales \\ Sydney, NSW 2052, Australia \\ \\
$^{b\,}$School of Mathematics and Statistics, University of Sydney \\ Sydney, NSW 2006, Australia \\ \\
$^{c\,}$Faculty of Mathematics and Information Science, Warsaw University of Technology \\ 00-661 Warszawa, Poland \\ }

\date{\vskip 45 pt \today \vskip 35 pt}

\begin{document}

\maketitle

\begin{abstract}
We study the upper and lower bounds for prices of European and American style options with the possibility of an external termination, meaning that the contract may be terminated at some random time. Under the assumption that the underlying market model is incomplete and frictionless, we obtain duality results linking the upper price of a vulnerable European option with the price of an American option whose exercise times are constrained to times at which the external termination can happen with a non-zero probability. Similarly, the upper and lower prices for an vulnerable American option are linked to the price of an American option and a game option, respectively. In particular, the minimizer of the game option is only allowed to stop at times which the external termination may occur with a non-zero probability.
\end{abstract}

\newpage
\tableofcontents
\newpage

\section{Introduction}    \label{sec1}

The goal of this work is to obtain duality results for the upper and lower price bounds of European and American style options with the possibility of an external termination, meaning that the contract may be terminated at some random time by an external force and, if this occurs, then the {\it promised payoff} $P$ is replaced by the {\it recovery payoff} $R$, which can be either higher or lower than $P$.

From the perspective of the financial mathematics literature, we would like to give an essential generalization of results from Szimayer \cite{S2005} who has studied unilateral valuation of vulnerable American options in an incomplete extension of the Black-Scholes model with a possibly unbounded risk-neutral default intensity under the hypothesis (H). In particular, our results furnish essential extensions of Theorem 3 in \cite{S2005}.  At first glance our results may appear similar to that of Szimayer \cite{S2005}. However, our methods of proofs are completely different from the approach in \cite{S2005} where the main arguments were based on the possibility of approximation in probability of any stopping time in the Brownian filtration by random times with unbounded intensities. In contrast, inspired by Fuhrman et al. \cite{FPZ2016}, we use a BSDE approach, which can be easily extended to a more general nonlinear setup, as studied, e.g., in \cite{ALR2022}. Also, we make minimal assumptions on the random time and move away from the standard assumptions, such as the existence of the default intensity, hypothesis (H), conditions {\bf (C)} and {\bf (A)}.
For the sake of generality, we will use optional stochastic calculus from \cite{G1981,G1982} and the recent results on a martingale representation from \cite{CDV2020,CY2022}.

To be more precise, by duality results we mean the link between the price bound computed using a family of martingale
measures and the issuer's superhedging price computed via reflected BSDE. Similar to \cite{S2005} we obtain the interpretation
of the upper or superhedging price of a vulnerable European option with the payoff $P$ and recovery $R$ in an incomplete model as the classical
arbitrage-free price in the underlying complete model of an American option with the payoff $P\I_{\llb T \rrb} + R\I_{\llb 0, T\llb}$ where $T$ is the terminal data. In contrast to the result obtained in Szimayer \cite{S2005}, a pertinent new feature in this work is that, rather than being able to exercise at any time up to the terminal date $T$, the potential exercise times of the American option is constrained to the right support of the (optional) hazard process. In the same vein, we also show that the upper and lower price bound of the vulnerable American option is an American option with payoff $(P\vee R\I_{\barS})\I_{\llb 0 , T\rrb}+P_T \I_{\llb T \rrb}$ and a game option where the reward of the maximizer and minimizer is given by $P$ and $R$, respectively. A study of the lower price for the vulnerable European option was deliberately omitted since it can be obtained from the corresponding result for a vulnerable American option.
The interesting new feature here is the dependence of the payoff on the support of the hazard process, which is related to $\barS$, and that two players in the zero-sum game do not share the same set of exercise times. In particular, the minimizer's exercise times are constrained to the right support of the optional hazard process. From a financial perspective, the support of the hazard process corresponds to times when external termination can occur with non-zero probability. The main message here is that the knowledge of the timings of the external termination can help improve the price bound of vulnerable options.

The current work also differs from the existing literature on a BSDE approach to the superhedging problem for defaultable instruments, as was studied, e.g., by Dumitrescu et al. \cite{DQS2017,DQS2018} in the case of a complete market and Grigorova et al. \cite{GQS2020,GQS2021} under market incompleteness. As we only examine the pricing problem up to a random time, we do not work directly with BSDEs in the progressively enlarged filtration but rather with the so-called {\it pre-default} or {\it reduced BSDE}. By doing so, we separate the default-free market, which is assume to be complete, from the market with default event, which is possibly incomplete. The advantage here is that rather than retrieving a BSDE with constrained jumps in the enlarged filtration, we obtain a more natural representation of the (pre-default) super-hedging price of a vulnerable European option through a related American option.

The structure of the paper is as follows. In Section \ref{sec2}, we first introduce the necessary tools for the study of random times from the general theory of stochastic processes and the theory of enlargement of filtration. We then give some recent auxiliary results from Choulli and Yansori \cite{CY2022} regarding the characterization of equivalent martingale measures in the progressive enlargement of a reference filtration with observations of a random time. By taking advantage of these results we illustrate how the Az\'ema supermartingale or the survival process, and hence the optional hazard process, associated with a random time can be computed under a change of equivalent martingale measure. To the best of our knowledge, the general results derived in the subsection are new and are crucial to the computation of the price bounds for vulnerable European and American options. For simplicity, we later assume that the optional hazard process is continuous (condition {\bf (A)} holds) and hence is equal to the hazard process. However, it must be stressed that the knowledge of the optional hazard is indispensable in studying the case where the optional hazard is purely discontinuous.

In Section \ref{sec3}, we consider the upper price bound of a vulnerable European option. By penalizing the hazard process and making use of results on generalized BSDEs from Li et al. \cite{LLR2}, we show in Theorem \ref{th3.1} that the upper price bound of the vulnerable European option coincides with the price of an American option with admissible exercise times constrained to the right support of the hazard process. In other words, the worst case scenario for the buyer of a vulnerable European option is essentially when the counterparty is able to select the timing of the default in an optimal way, in other words, an American option.

Before moving to the study of vulnerable American options in Section \ref{sec5}, we give in Section \ref{sec4} some auxiliary results on the Doob-Meyer-Mertens decomposition and the pre-default value of a vulnerable American option. In Section \ref{sec5}, we consider the upper and lower price of a vulnerable American option. We show in Theorem \ref{th6.1} that the upper price
of a vulnerable American option is formally equivalent to the price of an American option with a modified payoff. More interestingly, we show in Theorem \ref{th6.2} that the lower price of a vulnerable American option can be interpreted as the value process of a game option where the sets
of admissible exercise times are different for each of the two players. To be more specific, the minimizer is only allowed to exercise at stopping times that take values in the right support of the optional hazard process.

Regarding the background knowledge, for the general theory of stochastic processes, we refer to He et al. \cite{HWY1992} and the reader interested in stochastic calculus for optional semimartingales is referred to Gal'\v{c}uk \cite{G1982}. For more details on the theory of random times and enlargement of filtration with applications to problems arising in financial mathematics (such as credit risk modeling or insider trading), the interested reader may consult the monograph by Aksamit and Jeanblanc \cite{AJ2017} and the recent paper by Jeanblanc and~Li~\cite{JL2020}. For the reader's convenience, 
the required results on generalized BSDEs and generalized RBSDEs are reported in Li et al. \cite{LLR2}.

\section{Market Model and Optional Hazard Process}    \label{sec2}

We start by introducing the notation and recalling some fundamental concepts associated with modeling of a random time and the concept of the progressive enlargement of a reference filtration. We assume that a strictly positive and finite random time $\vt$, which is defined on a probability space $(\Omega, \cG, \PP)$, as well as some {\it reference filtration} $\FF$ are given. Then the enlarged filtration $\GG$ is defined as the {\it progressive enlargement} of $\FF$ by observations of $\vt$ (see, e.g., \cite{AJ2017}) and thus a random time $\vt$, which is not necessarily an $\FF$-stopping time, belongs to the set of all finite $\GG$-stopping times, which is denoted as $\STG$. We emphasize that the filtrations $\FF$ and $\GG$ are henceforth supposed to satisfy the usual conditions of $\PP$-completeness and right-continuity.

Recall that a stochastic process $X$ with sample paths possessing right-hand limits is said to be {\it $\FF$-strongly predictable} if it is $\FF$-predictable and the process $X_+$ is $\FF$-optional (Definition 1.1 in \cite{G1982}).
We will use the following notation for classes of processes adapted to the filtration $\FF$: \\
\vskip-10pt
\noindent  $\bullet$ $\cOff$, $\cPff$, $\ov{\mathcal{P}}(\FF)$ and $\cPrff$ are the classes of all real-valued, $\FF$-optional, $\FF$-predictable, $\FF$-strongly predictable and $\FF$-progressively measurable processes, respectively;\\
\noindent  $\bullet$ $\cOffd$, $\cPffd$, $\ov{\mathcal{P}}_d(\FF)$ and $\cPrffd$ are the classes of all $\RR^d$-valued, $\FF$-optional, $\FF$-predictable, $\FF$-strongly predictable and $\FF$-progressively measurable processes, respectively;\\
\noindent  $\bullet$ $\Martf$ (resp., $\Martfl$) is the class of all $\FF$-martingales (resp., $\FF$-local martingales);\\
\noindent  $\bullet$ $\Martft$ (resp., $\Martflt$) is the class of all $\FF$-martingales (resp., $\FF$-local martingales) stopped at $\vt $.

\noindent An analogous notation is used for various classes of $\GG$-adapted processes. For instance, $\cPgg$ denotes the class of all $\GG$-predictable processes, $\Martglt$ is the class of all $\GG$-local martingales, which are stopped at the random time $\vt $, etc.

In order to simplify the notation, we denote by $X\bigcdot Y$ the It\^o integral of $X$ with respect to a classical
(i.e., c\`adl\`ag) semimartingale $Y$, that is, $(X \bigcdot Y)_t := \int_{]0,t]} X_s\,dY_s$.
We will use the notation from Gal'\v{c}uk \cite{G1982} pertaining to a pathwise decomposition of a l\`adl\`ag process. If $C$ is an $\FF$-adapted, l\`adl\`ag process, then we write $C=C^c+C^d+C^g$ where the process $C^c$ is continuous, the c\`adl\`ag process $C^d$
equals $C^d_t:=\sum_{0\leq s\leq t}(C_{s}-C_{s-})$ and the c\`agl\`ad process $C^g$ is given by $C^g_t:=\sum_{0\leq s<t}(C_{s+}-C_s)$.
This also means that $C=C^r+C^g$ where the c\`adl\`ag process $C^r$ satisfies $C^r=C-C^g =C^c+C^d$. Notice that if $C$ is a c\`agl\`ad process,
then manifestly $C^d=0$ and thus $C^r=C^c$ is a continuous process. Similarly, if $C$ is a c\`adl\`ag process, then $C^g=0$ and
thus $C=C^r$. Furthermore, we write $\Delta C = \Delta C^d = C- C_{-}$ and $\Delta^+ C = \Delta C^g_+ = C_+ - C$  where $C^g_+$ is the c\`adl\`ag version of the c\`agl\`ad process $C^g$. Finally, we denote the stochastic exponential of an optional (l\`agl\`ad) semimartingale $X$ by $\mathcal{E}(X)$ and $\mathcal{E}_{s,t}(X)$ = $\mathcal{E}_t(X)/\mathcal{E}_s(X)$ for $s\leq t$ (see Theorem 5.1 in \cite{G1985}).

\subsection{Predictable and Optional Hazard Processes}  \label{sec2.1}

For a fixed random time $\vt$, we define the {\it indicator process} $A \in \cOgg$ by $A := \I_{\llb\vt,\infty\llb}$ so that $A_t=\I_\seq{\vt \leq t}$ for all $t\in \RR_+$. We denote by $\,^pA$ (resp., $\,^oA$) the $\FF$-predictable projection (resp., the $\FF$-optional projection) of $A$. Furthermore, we denote by $A^p$ (resp., $A^o$) the dual $\FF$-predictable projection (resp., the dual $\FF$-optional projection) of $A$. The BMO $\FF$-martingales $m$ and $n$ associated with $A^o$ and $A^p$, respectively, are defined as follows.

\bd \label{def2.1}
{\rm Let $m_t:=\EP[A^o_{\infty}\,|\,\cF_t]$ so $m_{\infty}=A^o_{\infty}$ and let $n_t:=\EP[A^p_{\infty}\,|\,\cF_t]$ so $n_{\infty}=A^p_{\infty}$.}
\ed

\noindent As in Az\'ema \cite{A1972}, we introduce the $\FF$-supermartingales $G$ and $\wtG$ associated with $\vt$.

\bd \label{def2.2}
{\rm The c\`adl\`ag process $G\in \cOff $, given by the equality $G_t:=\PP (\vt > t\,|\,\cF_t)$, is called the {\it Az\'ema supermartingale} of $\vt $ with respect to $\FF$.  The l\`adl\`ag process $\wtG \in \cOff$, given by the equality $\wtG_t:=\PP ( \vt \geq t\,|\,\cF_t)$, is called the {\it Az\'ema optional supermartingale} of $\vt$ with respect to~$\FF$.}
\ed

It is clear that $\wtG\geq G$ and the following equalities hold
\begin{equation} \label{eq2.1}
G={}^o(\I_{\llb 0,\vt\llb})={}^o(1-A),\quad \wtG={}^o(\I_{\llb 0,\vt\rrb})={}^o (1-A_{-}).
\end{equation}
The {\it $\FF$-predictable hazard process} of $\vt$, denoted as $\Gamma $, is given by (see Definition 1.6 in \cite{JL2020})
\begin{equation} \label{eq2.2}
\Gamma_t=\int_{]0,t]}G_{s-}^{-1}\,dA^p_s
\end{equation}
and the {\it $\FF$-optional hazard process} of $\vt$, denoted as $\wtGamma$, equals (see Definition 2.8 in \cite{JL2020})
\begin{equation} \label{eq2.3}
\wtGamma_t=\int_{[0,t]}\wtG_s^{-1}\,dA^o_s.
\end{equation}
Let us recall some well-known properties of the Az\'ema supermartingales $G$ and $\wtG$ and their relationship to the hazard processes
$\Gamma $ and $\wtGamma $ (see, e.g., Aksamit and Jeanblanc \cite{AJ2017}). In particular, for the proofs of assertions (iv)--(vi), we refer to Propositions 2.3 and 2.9 in Jeanblanc and Li \cite{JL2020} (see also Kardaras \cite{K2014,K2015}). It is worth noting that
$G$ is strictly positive if and only if $G_{-}$ is strictly positive or, equivalently, $\wtG$ is strictly positive.

\bl \label{lem2.1}
(i) We have that $G=n-A^p=m-A^o$ and $\wtG=m-A^o_{-}$ and thus
$$
G_t=\EP[A^p_{\infty }-A^p_{t}\,|\,\cF_t]=\EP[A^o_{\infty }-A^o_{t}\,|\,\cF_t],\quad \wtG_t=\EP[A^o_{\infty }-A^o_{t-}\,|\,\cF_t].
$$
(ii) The following equalities are valid: $\wtG_-=G_-$ and $\wtG_+=G_+=G$.\\
(iii) The following equalities are valid: $\wtG-G={}^o(\I_{\llb \vt \rrb})=\Delta A^o$ and $\wtG-G_{-}=\Delta m$.\\
(iv) If $G$ is strictly positive, then $G=\cE(-\wtGamma )\cE(\wt{N}) $ where $\wt{N}:=\wtG_{-}^{-1}\bcd m=G_{-}^{-1}\bcd m\in\Martfl$.\\
(v) If $\wtG$ is strictly positive, then $\wtG=\cE(-\wtGamma_{-})\cE(\wt{N})$ where $\wt{N}:=\wtG_{-}^{-1}\bcd m=G_{-}^{-1}\bcd m\in\Martfl$.\\
(vi) If $G$ is strictly positive, then $G= \cE(-\Gamma )\cE(N) $ where $N:=({}^p G)^{-1}\bcd n\in\Martfl$.
\el

The equality $G=n-A^p$  (resp., $\wtG=m-A^o_{-}$) gives the Doob-Meyer (resp., the Doob-Meyer-Mertens) decomposition in the
filtration $\FF$ of the supermartingale $G$ (resp., the optional supermartingale $\wtG$).
It is also known from the classical theory of enlargement of filtration that the $\GG$-martingale part of the Doob-Meyer
decomposition in the filtration $\GG$ of the bounded $\GG$-submartingale $A$, denoted by $\nG$, has the following representation
\begin{equation} \label{eq2.4}
\nG:=A-\I_{\rrb 0,\vt\rrb}G^{-1}_{-}\bcd A^p=A-\I_{\rrb 0,\vt\rrb}\bcd\Gamma=A-\Gamma^\vt.
\end{equation}
Furthermore, it was shown in Choulli et al. \cite{CDV2020} (see Theorem 2.3 in \cite{CDV2020} or Jeanblanc and Li \cite{JL2020}) that the following process $\mG$ is a $\GG$-martingale with integrable variation
\begin{equation} \label{eq2.5}
\mG:=A-\I_{\rrb 0,\vt\rrb}\wtG^{-1}\bcd A^o=A-\I_{\rrb 0,\vt\rrb}\bcd\wtGamma=A-\wtGamma^\vt.
\end{equation}
The processes $\nG$ and $\mG$ are known to belong to the class $\Martg$ but their properties are markedly different. In particular,
$\mG$ is a {\it pure default martingale} (see Definition 2.2 in Choulli et al. \cite{CDV2020}) whereas $\nG$ does not necessarily have
that property.

The following result is well known (see, e.g., Lemma 2.9 and Corollary 2.10 in \cite{AJ2017}).

\bl \label{lem2.2}
If $G$ is a strictly positive process, then for any bounded, $\FF$-predictable process $X$
\begin{equation} \label{eq2.6}
\EP(X_{\vt}\,|\,\cG_t)=\I_\seq{\vt \leq t}X_{\vt}+\I_\seq{\vt>t}G_t^{-1}\,\EP\Big[\int_{]t,\infty]}X_s\,dA^p_s\,\Big|\,\cF_t\Big]
\end{equation}
and for any bounded, $\FF$-optional process $X$
\begin{equation} \label{eq2.7}
\EP(X_{\vt}\,|\,\cG_t)=\I_\seq{\vt \leq t}X_{\vt}+\I_\seq{\vt>t}G_t^{-1}\,\EP\Big[\int_{]t,\infty]}X_s\,dA^o_s\,\Big|\,\cF_t\Big].
\end{equation}
\el

The following useful result is due to Aksamit et al. \cite{ACJ2015}.

\bp \label{pro2.1}
Let the $\FF$-stopping time $\wt \eta$ be given by $\wt \eta:=\inf \{t\in \RR_+:\wtG_{t-}>\wtG_{t}=0\}$.
If $M$ is an $\FF$-local martingale, then the process
\begin{equation} \label{eq2.8}
M^\vt-\I_{\llb 0,\vt\rrb}\wtG^{-1}\bigcdot [M,m]+\I_{\llb 0,\vt\rrb}\bigcdot\big(\Delta M_{\wt \eta}\I_{[\![\wt \eta,\infty [\![}\big)^p
\end{equation}
is a $\GG$-local martingale stopped at $\vt$.
\ep

In particular, if $\wtG$ is a strictly positive process, then for any $\FF$-local martingale $M$ we set
\begin{align} \label{eq2.9}
\sTr{M} & :=M-\wtG^{-1}\bigcdot [M,m] \qquad \mathrm{and} \qquad \cTr{M}: = \mathscr{T}(M^\vt)  =M^\vartheta-\wtG^{-1}\bigcdot [M,m]^\vartheta.
\end{align}
Note that $\cTr{M}$ is a $\GG$-local martingale stopped at $\vt$. If, in addition, all $\FF$-martingales are continuous (that is, if the so-called condition {\bf (C)} holds -- for instance, when $\FF$ is a Brownian filtration), then $\cOff =\cPff$ and thus the equalities $\wtG=G_-$ and $A^o=A^p$ are valid so that also $\mG=\nG$. Then equality \eqref{eq2.9} becomes 
\begin{equation} \label{eq2.10}
\sTr{M}=M-G^{-1}\bigcdot \langle M,n\rangle .
\end{equation}

\bl \label{lem2.3}
If $M$ is a uniformly integrable $\FF$-martingale, then the process
\begin{equation}   \label{eq2.11}
M^{\vt-}- G^{-1}\bcd [M,n]^{\vt-}
\end{equation}
is a $\GG$-local martingale.
\el

\begin{proof}
Let $H$ be a bounded $\GG$-predictable process and let $h$ be a bounded $\FF$-predictable process such that  $h\I_{\llb 0, \vt \rrb} = H\I_{\llb 0, \vt \rrb}$.
By using the dual $\FF$-predictable projection of $A$, we obtain
\begin{align*}
\EP[(H\bigcdot \I_{\llb 0,\vt\llb} M]_\infty)=\EP[(h \bigcdot M)_{\vt-}]=\EE[((h\bigcdot M)_{-}\bigcdot A^p)_\infty]= -\EP[((h\bigcdot M)_{-}\bigcdot G)_\infty].
\end{align*}
The integration by parts formula and Y\oe urp's lemma yield
\begin{align*}
 -\EP[((h\bigcdot M)_{-}\bigcdot G)_\infty]&=\EP[(G_-\bigcdot (h\bigcdot M))_\infty]+\EP[[h\bigcdot M,G]_\infty]\\
&=\EP[(h\bigcdot [M,n])_\infty]=\EP[(HG^{-1}\I_{\llb 0,\vt \llb}\bigcdot [M,n])_\infty]
\end{align*}
since $H = h$ on $\llb 0, \vt\rrb$ and it is known that $\{\vt > t\} \subset \{G_t> 0\}$.
\end{proof}

\subsection{Martingale Densities and Martingale Measures}  \label{sec2.0}

As in Choulli and Yansori \cite{CY2022}, we consider a market model composed of a finite family of underlying assets described by a semimartingale $S$ and complemented by an arbitrary random time $\vartheta$, which might not be an $\FF$-stopping time. The main goal in \cite{CY2022} was to obtain an explicit description
of the set of all deflators for the stopped process $S^{\vt}$ and to analyze the No Free Lunch with Vanishing Risk (NFLVR) property for the stopped market model $(S^{\vt},\GG ,\PP)$. As was pointed out in \cite{CY2022}, the class of all deflators constitutes the dual set of all admissible wealth processes for the market $(S^{\vt},\GG ,\PP)$ and hence it can be used to address the hedging and pricing problems for derivative securities using, in particular, previous results on the classification of martingales in filtration $\GG$ and representation
theorems for a progressively enlarged filtration obtained in Choulli et al. \cite{CDV2020}.

We refer the reader to \cite{CDV2020} and \cite{CY2022} for details and we only present some results from \cite{CY2022} that will be used in what follows. In particular, the reader is referred to Definition 2.1 in \cite{CY2022} for the concepts of a {\it deflator} and to Definition 4.1 therein for the NFLVR property. To formulate our assumptions, we need to recall a generic definition of a martingale deflator and martingale density where $\HH$ is any filtration (in our setup, $\HH$ can be taken to be either $\FF$ or $\GG$) and $X$ is an arbitrary $\HH$-semimartingale.

\bd
Consider a triplet $(X,\HH ,\PP)$ where $X$ is an $(\HH ,\PP)$-semimartingale. \\
(i) A stochastic process $Z$ is a {\it local martingale deflator} for $(X,\HH ,\PP)$ if $Z_0=1,\, Z>0$ and there exists a real-valued, $\HH$-predictable process $\varphi$ such that $0<\varphi\leq 1$ and the processes $Z$ and $Z (\varphi \bigcdot X)$ are $(\HH ,\PP)$-local martingales.
The class of local martingale densities for $(X,\HH ,\PP )$ is denoted by $\cZ_{loc}(X,\HH ,\PP)$. \\
(ii) A stochastic process $Z$ is a {\it martingale density} for $(X,\HH ,\PP)$ if $Z$ belongs to $\cZ_{loc}(X,\HH ,\PP)$ and is
a uniformly integrable $(\HH ,\PP)$-martingale. The class of martingale densities for $(X,\HH ,\PP )$ is denoted by $\cZ(X,\HH ,\PP)$.
\ed

It is known that a market model $(X,\HH,\PP)$ enjoys the NFLVR property (see, e.g., Definition 4.1 in \cite{CY2022}) if and only if the class $\cZ (X,\HH,\PP)$ of martingale densities for $(X,\HH ,\PP)$ is nonempty.

We consider a market model with a finite horizon date $T>0$.
Let $S$ be a $d$-dimensional $\FF$-semimartingale, which is defined on a probability space $(\Omega, \cG, \PP)$. As customary,
we assume that $S$ represents discounted prices of non-dividend-paying assets with prices $S^1,S^2,\dots, S^d$.
As in  Choulli and Yansori \cite{CY2022}, we first consider the market model $(S,\FF ,\PP)$ and subsequently examine its modification represented by the triplet $(S^{\vt},\GG ,\PP)$ where $\vt$ is a fixed random time and
$\GG$ is the progressive enlargement of $\FF$ with $\vt$. Since we consider the case of a finite horizon date $T$ we also
have that $S^{\vt}=S^{\vt\wedge T}$. We will henceforth assume that the class $\cZ (S,\FF,\PP)$ is nonempty, which is equivalent to the NFLVR property of the market model $(S,\FF ,\PP)$.

Theorem 3.2 in Choulli and Yansori \cite{CY2022} describes the class of all local martingale deflators for the market $(S^{\vt},\GG,\PP)$
and Theorem 4.2 studies martingale densities for $(S^{\vt},\GG,\PP)$ when $G>0$. For the reader's convenience, we state some of their
findings in Theorem \ref{thmxx}. As in \cite{CY2022}, we denote
\[
\mathcal{I}^{o,\FF}_{loc}(m^{\GG},\GG) := \big\{ K \in \cOff :|K|G \wtG^{-1} \I_{\{\wtG>0\}} \bigcdot A \in \cA_{loc}(\GG,\PP) \big\}
\]
where $\cA_{loc}(\GG)$ is the set of all $\GG$-adapted, c\`adl\`ag  with locally integrable variation under $\PP$
It is known (see Theorem 2.3 (b) in \cite{CY2022}) that for every process $K \in \mathcal{I}^{o,\FF}_{loc}(m^{\GG},\GG)$ the
process $K \bigcdot m^{\GG}$ is a $\GG$-local martingale with locally integrable variation.
Furthermore, we denote by $L^1_{loc}(\cPrff, \PP \otimes dA)$ the class of all $\FF$-progressively measurable processes
locally integrable with respect to $\PP \otimes dA$. The following result can be deduced from Theorems 3.2 and 4.2 in Choulli and Yansori \cite{CY2022}.

\bt  \label{thmxx}
Suppose that $G>0$ (and hence also $\wtG>0$). Then: \\
(i) If the market model $(S,\FF ,\PP)$ satisfies NFLVR, then the market model $(S^{\vt},\GG,\PP)$ also satisfies NFLVR.
Equivalently, if the class $\cZ(S,\FF ,\PP)$ is nonempty, then the class $\cZ(S^{\vt},\GG,\PP)$ is nonempty.
Furthermore, if a process $Z^{\FF}$ belongs to $\cZ(S,\FF ,\PP)$, then the process $Z^{\GG} := (Z^{\FF})^{T\wedge \vt} / \cE (G^{-1}_{-} \bigcdot m)^{T\wedge \vt}$ belongs to $\cZ(S^{\vt},\GG,\PP)$. \\
\noindent (ii) Let the couple $(\varphi^{(o)},\varphi^{(pr)})\in \mathcal{I}^{o,\FF}_{loc}(m^{\GG},\GG) \times L^1_{loc}(\cPrff, \PP \,\otimes\, dA )$ be bounded processes such that $\EP \big[\varphi^{(pr)}_{\vt}\,|\,\cF_{\vt}\big]=0$ and
\begin{align} \label{xposit1}
\varphi^{(pr)}>-1,\qquad \varphi^{(o)}>-\wtG{G}^{-1}, \ \PP \otimes dA\mbox{\rm-a.e.}
\qquad \varphi^{(o)}(\wtG -G) < \wtG , \ \PP \otimes \Delta A^o\mbox{\rm-a.e.}
\end{align}
and let $Z^{\FF}$ be any martingale density for the market model $\cZ(S,\FF ,\PP)$. If the process $(\eta^{\varphi}_t,\, t \in [0,T])$ given by
\begin{align} \label{xposit2}
\eta^{\varphi}:=\frac{(Z^{\FF})^{\vt}}{\cE (G^{-1}_{-} \bigcdot m)^{\vt}}\,\cE\big(\varphi^{(o)}\bcd \mG \big)\cE\big(\varphi^{(pr)}\bcd A \big)
\end{align}
is a uniformly integrable $(\GG,\PP)$-martingale, then it belongs to the class $\cZ(S^{\vt},\GG,\PP)$,
that is, the process $\eta^{\varphi}$ is a martingale density for the market model $\cZ(S^{\vt},\GG,\PP)$.
\et

\brem \label{xrem2.1}
Notice that conditions \eqref{xposit1} are necessary and sufficient for the property of strict positivity of the
process $\eta^{\varphi}$ given by \eqref{xposit2}. This is readily seen for the first inequality in \eqref{xposit1},
which is equivalent to the property $\cE\big(\varphi^{(pr)}\bcd A \big)>0$.  Similarly, the condition $\varphi^{(o)}>-\wtG{G}^{-1}$ is needed
to ensure that $\cE\big(\varphi^{(o)}\bcd \mG \big)$ is strictly positive on the graph $[\![\vt]\!]$ of $\vt$
and the inequality $\varphi^{(o)}(\wtG -G) < \wtG$ is equivalent to the strict positivity of $\cE\big(\varphi^{(o)}\bcd \mG \big)$
on $[\![0,\vt [\![$. To check these conditions, it suffices to use the explicit representation for the Dol\'eans exponential,
equation \eqref{eq2.5} and the equalities $\wtG -G = \Delta A^o = \wtG \Delta \wtGamma $, which follow from Lemma \ref{lem2.1} (i)
and equation \eqref{eq2.3}. In the special case where $A^o$ (or, equivalently, $\wtGamma$) is a continuous process, we have that $\wtG = G$ and thus conditions  \eqref{xposit1} become: $\varphi^{(pr)}>-1,\, \varphi^{(o)}>-1, \, \PP \otimes dA\mbox{\rm-a.e.}$.
\erem

\brem \label{xrem2.1x}
In general, the process $\eta^{\varphi}$ given by \eqref{xposit1} is a local martingale under $\PP$ and thus it is a local martingale deflator
for $\cZ(S^{\vt},\GG,\PP)$. However, we postulate that $\eta^{\varphi}$ is a martingale under $\PP$ and thus it is a martingale density for
for the market model $\cZ(S^{\vt},\GG,\PP)$.  It is shown in Theorem 4.2 in \cite{CY2022} that if $\varphi^{(o)}\geq 0$, then $\eta^{\varphi}$ is a martingale under $\PP$ and hence is a martingale density for $(S^{\vt},\GG,\PP)$  but the inequality $\varphi^{(o)}\geq 0$ is too restrictive if one wishes to describe the class of all martingale densities for $(S^{\vt},\GG,\PP)$.
\erem

The condition $\varphi^{(o)}\geq 0$ is too restrictive and is only a sufficient condition used in Theorem 4.2 of \cite{CY2022} to ensure that $\eta^{\varphi}$ is a uniformly integrable $\GG$ martingale. We present the following lemma to give sufficient conditions for that property to hold. In view of Lemma \ref{lem3.xx}, we will later postulate in Assumption \ref{ass3.1} that the hazard process $\Gamma$ is continuous and satisfies suitable integrability conditions.

\bl \label{lem3.xx}
Assume that $\varphi^{(o)} + G^{-1}\wtG > 0$, the continuous part of the optional hazard process $\wtGamma^c$ satisfies $\EE[e^{\wt\Gamma_{T\wedge\vartheta}^c}]< \infty$ and the number of jumps of $\wtGamma$ is bounded.
Then the process $\eta^{\varphi}$ is a uniformly integrable $(\GG,\PP)$-martingale.
\el

\begin{proof}
We observe (see also equation (4.80) in \cite{CY2022}) that
\begin{align*}
0 	& \leq \cE\big(\varphi^{(o)}\bcd \mG \big) \leq \cE\big(-\varphi^{(o)} \I_{{\rrb 0, \vartheta \llb}} \bcd \wtGamma \,\big)\big(1+ G_\vartheta\wtG^{-1}_\vartheta \varphi^{(o)}_\vartheta \I_{\llb \vartheta,\infty \rrb\}} \big).
\end{align*}
From the fact that $\wtG \geq G$ and that $\varphi^{(o)}$ is bounded, one can deduce that the second term in the product above is bounded. The first term on the other hand can be estimated as follows
\begin{align*}
\cE\big( -\varphi^{(o)} \I_{{\rrb 0, \vartheta \llb}} \bcd \wtGamma\big)
	& = \exp\big( -(\wtG G^{-1} + \varphi^{(o)}) \I_{{\rrb 0, \vartheta \llb}} \bcd \wtGamma^c \big) \exp\big( \wtG G^{-1}\I_{{\rrb 0, \vartheta \llb}} \bcd \wtGamma^c \big)   \prod_{s\leq \cdot } \big( 1-\varphi^{(o)}_s\Delta \wtGamma_s\big)\\
	& \leq \exp\big( \wtGamma^c \big)   \prod_{s\leq \cdot } \big( 1+ \Delta \wtGamma_s\big) \leq \exp\big( \wtGamma^c \big)   2^N
\end{align*}
where in the last inequality we have used the assumption that $\varphi^{(o)} + G^{-1}\wtG> 0$ and the number of jumps is bounded. From the above we obtain
\begin{gather*}
\EE\Big[\sup_{t\leq T}|\cE_t\big(\varphi^{(o)}\bcd \mG \big)|\Big] \leq \EE\big[\exp\big( \wtGamma^c_{T\wedge\vartheta} \big)\big]   2^N < \infty
\end{gather*}
and hence $\cE\big(\varphi^{(o)}\bcd \mG \big)$ is a uniformly integrable $(\GG,\PP)$-martingale.
\end{proof}

We henceforth work under the postulate that the class $\cZ (S,\FF,\PP)$ has a unique element, which is denoted by $Z^{\FF}$. Then the probability measure $\QQ$ on $(\Omega ,{\cal F}_T)$ given by $d\QQ/d{\PP}=Z^{\FF}_T$ is the unique equivalent local martingale measure (ELMM) for the model $(S,\FF ,\PP)$, in the sense that there exists a real-valued, $\FF$-predictable process $\varphi$ such that $0<\varphi\leq 1$ and the process $\varphi \bigcdot S$ is an $(\FF ,\QQ)$-local martingale (i.e., the process $S$ is a $\sigma$-martingale under $\QQ$). We work throughout under the standing assumptions of the strict positivity of $G$ and completeness of the market model $\cZ(S,\FF ,\PP)$.

\bhyp \label{ass2.1} We postulate that:\\
(i) the Az\'ema supermartingale $G$ of $\vt $ under $\PP$ (and hence $G_{-}$ and $\wtG$) are strictly positive; \\
(ii) the class $\cZ(S,\FF ,\PP)$ is nonempty and has a unique element $Z^{\FF}$, which defines a unique ELMM $\QQ$
for the complete market model $(S,\FF ,\PP)$; \\
(iii) there exists an $\RR^{d}$-valued, $(\FF,\PP)$-local martingale $M$ with the predictable representation property for the filtration $\FF$.
\ehyp

In view of Assumption \ref{ass2.1} (ii), the market model $\cZ(S,\FF ,\PP)$ has a unique ELMM $\QQ$.
The $(\FF,\QQ)$-martingale associated with $M$ through the Girsanov theorem is denoted by $\whMM$ or more explicitly,
\begin{gather}
M^\QQ = M - \frac{1}{Z^\FF_-}\bcd[M, Z^\FF]. \label{MQ}
\end{gather}
It is known that $\whMM$ the predictable representation property holds in the filtration $\FF$ under $\QQ$ (see Aksamit and Fontana \cite{AF2019}).

\bl \label{lmbbb}
There exists a $\FF$-predictable process $\varphi^{(p)}$ such that
\begin{align} \label{xposit3x}
\frac{Z^{\FF}_{T\wedge \vt}}{\cE_{T\wedge \vt} (G^{-1}_{-} \bigcdot m)}=\cE_{T}\big(\varphi^{(p)}\bcd \cTr{M} \big)
\end{align}
where the $(\GG ,\QQ)$-martingale $\cTr{M}$ is given by equation \eqref{eq2.9}.
\el

\begin{proof}
The asserted equality can be deduced from Assumption \ref{ass2.1} (iii). We may represent the strictly positive $(\FF,\PP)$-martingale
 $Z^{\FF}$ as $Z^{\FF}= \cE(M^\FF)$ where $M^\FF$ is some $(\FF,\PP)$-martingale. From equation (3.33) in Choulli
and Yansori \cite{CY2022} we know that
\begin{align*}
\frac{(Z^{\FF})^{\vt}}{\big(\cE(G^{-1}_{-} \bigcdot m)\big)^{\vt}} = \cE\big( \cTr{M^\FF} - G^{-1}_-\bcd \cTr{m}  \big).
\end{align*}
By Assumption \ref{ass2.1} (iii), the martingale representation holds for $M$ and thus it is clear from \eqref{eq2.9} that $\cTr{M^\FF}  = \zeta \bcd \cTr{M} $
and $\cTr{m} = \psi \bcd \cTr{M}$ for some $\FF$-predictable processes $\zeta$ and $\psi$. Then
\begin{align*}
 \cE\big( \cTr{M^\FF} - G^{-1}_-\bcd \cTr{m}  \big)
  = \cE\big( (\zeta- \psi G^{-1}_-)\bcd \cTr{M}  \big) = \cE\big(\varphi^{(p)}\bcd \cTr{M}  \big),
\end{align*}
which gives the desired equality.
\end{proof}

In view of Assumption \ref{ass2.1}, we henceforth assume that the process $\varphi^{(p)}$ is fixed throughout and we introduce the following
definition.

\bd
We denote by $\Phi$ the class of all processes $\varphi= (\varphi^{(o)},\varphi^{(pr)})$ satisfying the assumptions of Theorem \ref{thmxx} (ii) and
such that the process $\eta^{\varphi}$, which is given by
\begin{align} \label{xposit4}
\eta^{\varphi}=\cE\big(\varphi^{(p)}\bcd \cTr{M} \big)\cE\big(\varphi^{(o)}\bcd \mG \big)\cE\big(\varphi^{(pr)}\bcd A \big)= \eta^{\varphi^{(p)}}\eta^{\varphi^{(o)}}\eta^{\varphi^{(pr)}} ,
\end{align}
is a uniformly integrable $(\GG,\PP)$-martingale.
\ed

Observe that for any $\varphi \in \Phi$ the process $\eta^{\varphi}$ belongs to $\cZ(S^{\vt},\GG,\PP)$ and
thus it defines a probability measure $\QQ^{\varphi}$ on $(\Omega , \cG_T)$ through the equality $d\QQ^{\varphi}/d\PP = \eta^{\varphi}_T$
or, equivalently,
\begin{align} \label{xposit5}
\frac{d\QQ^{\varphi}}{d\PP}=\cE_{T}\big(\varphi^{(p)}\bcd \cTr{M} \big) \cE_T\big(\varphi^{(o)}\bcd \mG \big)\cE_T\big(\varphi^{(pr)}\bcd A \big)
=\eta^{\varphi^{(p)}}_{T} \eta^{\varphi^{(o)}}_T\eta^{\varphi^{(pr)}}_T = \eta^{\varphi}_T.
\end{align}
It follows from Theorem \ref{thmxx} (ii) that any probability measure $\QQ^{\varphi}$ is an ELMM for $(S^{\vt},\GG,\PP)$, which entails the incompleteness of the market model $(S^{\vt},\GG,\PP)$ since it is clear that the uniqueness of an ELMM does not hold.

\subsection{Optional Hazard Process under a Martingale Measure}      \label{sec2.2}

We are now in a position to start a study of arbitrage-free pricing of defaultable claims of either a European or an American style in the market model $(S^{\vt},\GG,\PP)$. Our first goal is to compute the $\FF$-optional projection and the dual $\FF$-optional projection of the process $A$ after an equivalent change of a probability measure. For brevity, given two equivalent probability measures on $(\Omega,\cG_T)$,
denoted as $\PP$ and $\QQphi$, we shall only explicitly indicate the probability measure under which the relevant processes are either
defined or computed when we work under $\QQphi\in\cQ$. Hence, for instance, we denote by $G$ the $\FF$-optional projection of $1-A$ under $\PP$, while we write $G^{\QQphi}$ (or, simply, $G^{\varphi}$) for the $\FF$-optional projection of $1-A$ under $\QQphi$. For the ease of notation, this convention will be applied to all processes throughout the paper, that is, the superscript $\QQphi$ will be simplified to $\varphi$.

As was already mentioned, the process $\mG=A-\wtGamma^\vt$ given by \eqref{eq2.5} is a pure default $\GG$-martingale of finite variation and $\mG=(\mG)^{\vt}$. In addition, for any bounded $\FF$-optional process $K$ we have
\begin{align} \label{eq2.12}
\EP[(K \bcd (\mG)^o)_\infty]=\EP[(K\bcd \mG)_\infty]=0
\end{align}
since the process $K \bcd \mG$ is a $\GG$-martingale. The key message from \eqref{eq2.12} is that $(\mG)^o=0$ or, equivalently,
$A^o=(\wtGamma^\vt)^o$.

Notice that the equality $A^o=(\wtGamma^\vt)^o$ can be checked directly as follows: $\wtGamma^\vt=(1-A_-)\wtG^{-1}\bcd A^o$ and thus
$(\wtGamma^\vt)^o=\,^o(1-A_-)\wtG^{-1}\bcd A^o=A^o$. Equality \eqref{eq2.12} shows that in order to compute the dual $\FF$-optional
projection of $A$ under an equivalent probability measure $\QQphi\in\cQ$, it suffices to identify the optional Jeulin-Yor formula
(see, e.g., Lemma 2.7 in \cite{JL2020}) for the pure default martingale $\mG$ under $\QQphi$.

The following lemma is used in the proof of Proposition \ref{pro2.2}.

\bl \label{lem2.4}
Let $(\whtau_n)_{n\in\NN}$ be a sequence of $\GG$-stopping times such that $\lim_{\, n\rightarrow \infty}\whtau_n=\infty$. If $(\tau_n)_{n\geq 1}$ is a sequence of $\FF$-stopping times such that $\whtau_n\wedge\vt = \tau_n\wedge\vt$ for all $n\in\NN$, then $\tau :=\lim_{\,n\rightarrow \infty}\tau_n\geq\nu:=\inf\{t\in\RR_+:G_t=0\}$.
\el

\begin{proof}
It is clear that $\{\whtau_n \wedge \vt>t\}=\{\tau_n\wedge\vt>t\}$ for all $t\geq 0$. By sending $n$ to infinity and using the monotone convergence theorem, we obtain the equality $\PP (\vt > t\,|\,\cF_t)=\PP(\vt > t\,|\,\cF_t)\I_{\{\tau>t\}}$ and hence $\tau>\nu_k$ for all $k$ where $\nu_k=\inf\,\{t\in\RR_+:G_t\leq 1/k\}$. We conclude that $\I_{\cap_k\rrb\nu_k,\infty\llb}(\tau)=1$ and thus the equality $\tau\geq \nu$ is established. It is obvious that $\tau=\nu=\infty$ if Assumption \ref{ass2.1} is satisfied.
\end{proof}

\bp \label{pro2.2}
Let Assumption \ref{ass2.1} be satisfied and let
\begin{align*}
\oLambda  :=(1+\varphi^{(o)}(1-\Delta\wtGamma))\bcd\wtGamma  \quad \textrm{and} \quad \omG& :=A-\oLambda^\vt.
\end{align*}
Then the following statements hold for every $\varphi \in \Phi$:\\
(i) if $K$ is a bounded, $\GG$-optional process, then $K\bcd\omG$ is a $\GG$-local martingale under $\QQphi\in\cQ$; \\
(ii) the $\FF$-optional hazard process of $\vt$ under $\QQphi\in\cQ$ satisfies
\begin{align} \label{eq2.13}
\oGamma = \wtGaphi :=(\wt{G}^{\varphi})^{-1}\bcd A^{o,\QQphi}=\big(1+\varphi^{(o)}(1-\Delta \wtGamma)\big)\bcd\wtGamma=\oLambda.
\end{align}
\ep

\begin{proof}
(i) Since the martingales $\varphi^{(p)}\bcd \cTr{M},\, \varphi^{(o)}\bcd m^\GG$ and $\varphi^{(pr)}\bcd A$ are orthogonal to each other, it suffices to show that the process $\eta^{\varphi^{(o)}}(K\bcd\omG)$ is a $\GG$-local martingale under $\PP$. By the It\^o formula
\begin{align*}
& d(\eta^{\varphi^{(o)}}_t(K\bcd\omG)_t)\\
& =(K\bcd\omG)_{t-}\,d\eta^{\varphi^{(o)}}_t+\eta^{\varphi^{(o)}}_{t-}K_t\,d\omG_t+K_t\,d[\eta^{\varphi^{(o)}},\omG]_t\\
&=(K\bcd\omG)_{t-}\,d\eta^{\varphi^{(o)}}_t+\eta^{\varphi^{(o)}}_{t-}K_t\,d(A_t-\wtGamma_t^\vt)
+\eta^{\varphi^{(o)}}_{t-}K_t\,d(\wtGamma_t-\oLambda_t)^\vt+\eta^{\varphi^{(o)}}_{t-}\varphi_t^{(o)} K_t\,d[\mG,\omG]_t
\end{align*}
where the last term can be made more explicit
\begin{gather*}
\eta^{\varphi^{(o)}}_{t-}\varphi_t^{(o)}K_t\,d[\mG,\omG]_t = \eta^{\varphi^{(o)}}_{t-}\varphi_t^{(o)}K_t\big((1-\Delta\wtGamma_t-\Delta \oLambda_t)\,dA_t+\Delta \wtGamma_t\,d\oLambda_t\big).
\end{gather*}

Consequently,
\begin{align*}
 d(\eta^{\varphi^{(o)}}(K\bcd\omG)_t)
& = (K\bcd\omG)_{t-}\,d\eta^{\varphi^{(o)}}_t+\eta^{\varphi^{(o)}}_{t-}K_t\,d(A_t-\wtGamma_t^\vt)
  +\eta^{\varphi^{(o)}}_{t-}K_t\,d(\wtGamma_t-\oLambda_t)^\vt \\
 &\quad  +\eta^{\varphi^{(o)}}_{t-}\varphi^{(o)}_tK_t\big((1-\Delta\wtGamma_t-\Delta \oLambda_t)\,dA_t+\Delta \wtGamma_t\,d \oLambda_t\big)\\
&=(K\bcd\omG)_{t-}\,d\eta^{\varphi^{(o)}}_t+ \eta^{\varphi^{(o)}}_{t-} K_t\,d\mG_t+\eta^{\varphi^{(o)}}_{t-}\varphi_t^{(o)} K_t(1-\Delta \wtGamma_t-\Delta\oLambda_t)\,d\mG_t+dC_t
\end{align*}
where the last term satisfies
\begin{align*}
dC_t&=\eta^{\varphi^{(o)}}_{t-}K_t \big( \varphi^{(o)}_t(1-\Delta\wtGamma_t-\Delta\oLambda_t)\,d\wtGamma^\vt_t+\varphi_t^{(o)} \Delta \wtGamma_t \,d\oLambda_t+d(\wtGamma_t-\oLambda_t)^\vt \big) \\
&=\eta^{\varphi^{(o)}}_{t-}K_t \big((1+\varphi^{(o)}_t(1-\Delta \wtGamma_t))\,d\wtGamma_t^\vt-d\oLambda_t^\vt\big)=0
\end{align*}
since $\oLambda:=(1+\varphi^{(o)}(1-\Delta \wtGamma))\bcd\wtGamma$. Hence $\eta^{\varphi^{(o)}}(K\bcd \omG)$ is a $\GG$-local martingale under $\PP$. Notice that we have not identified the hazard process of $\vt$ under $\QQphi$ since it is not yet known whether $\oLambda$ is equal to
$(\wt{G}^{\varphi})^{-1}\bcd A^{o,\QQphi}$ (see equation \eqref{eq2.3}).

\noindent (ii) In order to find $A^{o,\QQphi}$, it suffices to repeat the computations leading to equation \eqref{eq2.12} but working under $\QQphi$. However, the situation is slightly more complicated here since we only know from part (i) that $\omG$ is a $\GG$-local martingale under $\QQphi$ and thus we need to use the method of localization and Lemma \ref{lem2.4}. Let us thus assume that $(\whtau_n)_{n\in\NN}$ is a $\GG$-localizing sequence for $\omG$ and $(\tau_n)_{n\in\NN}$ is the corresponding $\FF$-reduction so that the equality $\whtau_n \wedge \vt = \tau_n \wedge \vt$ holds for every $n\in\NN$ and let $K$ be a bounded, $\FF$-optional process. From part (i),
we know that $K\bcd (\omG)^{\whtau_n}$ is a $\GG$-martingale under $\QQphi$ and thus
\begin{align*}
0=\EQphi \big[(K\bcd (\omG)^{\whtau_n})_\infty\big]=\EQphi \big[(K\I_{\llb 0,\tau_n \rrb}\bcd (\omG)^\vt)_\infty\big]=
\EQphi \big[(K\I_{\llb 0,\tau_n\rrb}\bcd(\omG)^{o,\QQphi})_\infty\big],
\end{align*}
which implies that the equality $(\omG)^{o,\QQphi}=0$ holds on the interval $\llb 0,\tau_n\rrb$ for every $n\in\NN$.
Since $\omG:=A-\oLambda^\vt$, this means that the dual $\FF$-optional projection of $A$
under $\QQphi$ coincides, on the interval $\llb 0,\tau_n\rrb$ for every $n\in\NN$, with the dual $\FF$-optional projection of $\oLambda^{\vt}$ under $\QQphi$, that is, the equality $A^{o,\QQphi}=(\oLambda^\vt)^{o,\QQphi}$ holds on the stochastic interval $\cup_n\llb 0,\tau_n\rrb$.
Since $G$ is assumed to be strictly positive we deduce from Lemma \ref{lem2.4} that $\tau=\lim_{\,n\rightarrow \infty}\tau_n=\infty$ and thus
\begin{align} \label{eq2.14}
A^{o,\QQphi} = (\oLambda^\vt)^{o,\QQphi} = \wtGphi  \oLambda ,
\end{align}
which in turn implies that $\wtGaphi :=(\wtGphi)^{-1} \bcd A^{o,\QQphi} = \oLambda$ and thus the proof is completed.
\end{proof}

Observe that $G^{\varphi} >0$ and hence also $\wtGphi >0$ so that, in view of assertions (iv) and (v) in Lemma \ref{lem2.1}, they have under $\QQphi$ the multiplicative decompositions $G^{\varphi} =\cE(-\wtGaphi)\cE(\wt N^{\varphi})$ and $\wtGphi=\cE(-\wtGaphi_-) \cE(\wt N^{\varphi})$. Recall also from Lemma \ref{lem2.1} (iii) that $G=\cE(-\wtGamma)\cE(\wt N)$ where $\wt N = G_{-}^{-1}\bcd m$. The next result shows that the $\FF$-local martingale $\cE(\wt N^{\varphi})$ under $\QQphi$ can be represented in terms of the processes $\cE(\wt N)$ and $\eta^\varphi$.

\bl  \label{lem2.5}
The $\FF$-local martingale $\cE(\wt N^{\varphi})$ in the multiplicative decomposition of $G^{\QQphi}$ satisfies
\begin{align*}
\,^o(\eta^\varphi)\cE(\wt N^{\varphi}) & = \cE(\wt N)\cE(\varphi^{(p)}\bcd \sTr{M}) = \cE(G_{-}^{-1}\bcd m+ \varphi^{(p)}\bcd M).
\end{align*}
\el

\begin{proof}
Using equation \eqref{eq2.1} under $\QQphi$ and the abstract Bayes formula, we obtain
\begin{align*}
G^{\varphi}=\,^{o,\QQphi}(1-A)=\,^{o,\QQphi}\big(\I_{\llb 0,\vt \llb}\big)= \,^{o}\big(\eta^\varphi\I_{\llb 0,\vt \llb}\big)/\,^o(\eta^\varphi).
\end{align*}
On the interval $\llb 0,\vt \llb$, we have $\eta^{\varphi^{(pr)}} = 1$ and $\eta^{\varphi^{(p)}} = \cE(\varphi^{(p)}\bcd \wt M)$, which are $\FF$-optional processes.
We also note that $\varphi^{(o)}\bcd\mG=\varphi^{(o)}_\vt \I_{\llb \vt,\infty \llb}-\varphi^{(o)}\bcd\wtGamma^\vt$ so that
\begin{align*}
\eta^\varphi \I_{\llb 0,\vt\llb}=\cE(\varphi^{(p)}\bcd \cTr{M})\cE(\varphi^{(o)}\bcd\mG)\I_{\llb 0,\vt\llb}=\cE(\varphi^{(p)}\bcd \sTr{M})\cE(-\varphi^{(o)}\bcd\wtGamma)\I_{\llb 0,\vt\llb}.
\end{align*}

By taking the $\FF$-optional projection under $\PP$ and using Lemma \ref{lem2.1} (vi), we obtain
\begin{align*}
\,^{o}(\eta^\varphi\I_{\llb 0,\vt\llb}) & =\,^o(\cE(\varphi^{(p)}\bcd \sTr{M})\cE(-\varphi^{(o)}\bcd\wtGamma)\I_{\llb 0,\vt\llb}) =\cE(\varphi^{(p)}\bcd \sTr{M})\cE(\wt N)\cE(-\varphi^{(o)}\bcd\wtGamma)\cE(-\wtGamma)\\
& =\cE(\varphi^{(p)}\bcd M + G^{-1}_-\bcd m)\cE(-(\varphi^{(o)}(1-\Delta\wtGamma)+1)\bcd\wtGamma)
 =\cE(\varphi^{(p)}\bcd M + G^{-1}_-\bcd m)\cE(-\wtGaphi)
\end{align*}
where we have used the equality $\cE(X)\cE(Y)=\cE(X+Y+[X,Y])$ and the fact that
\begin{align*}
& \cE\big(\varphi^{(p)}\bcd \sTr{M}+G^{-1}_-\bcd m +\varphi^{(p)}G^{-1}_-\bcd[\sTr{M},m]\big) \\
&=\cE\big(\varphi^{(p)}\bcd M-\varphi^{(p)}\wtG^{-1}\bcd [M,m]+G^{-1}_-\bcd m+\varphi^{(p)}G^{-1}_-\bcd[M,m]-\varphi^{(p)}\wtG^{-1} G^{-1}_-\bcd[[M,m],m]\big) \\
&=\cE\big(\varphi^{(p)}\bcd M+G^{-1}_-\bcd m \big).
\end{align*}
The last equality above holds since $\wtG - G_-=\Delta m$ and $\Delta m\bcd [M, m]=[[M,m],m]$ so that
\begin{align*}
-\wtG^{-1}\bcd [M,m]+G^{-1}_-\bcd [M, m]-\wtG^{-1}G^{-1}_{-}\bcd [[M,m],m]=0.
\end{align*}
Finally, since $G^{\varphi}=\cE(-\wtGamma^{\varphi})\cE(\wt N^{\varphi})$ we have
$\cE(\wt N^{\varphi})=\cE\big(\varphi^{(p)}\bcd M+G^{-1}_-\bcd m \big)/\,^o(\eta^\varphi)$ and from the abstract Bayes formula, it is clear that the process $\cE\big(\varphi^{(p)}\bcd M+G^{-1}_-\bcd m \big)/\,^o(\eta^\varphi)$ is an $\FF$-local martingale
under $\QQphi$.
\end{proof}

\bl \label{nlemm}
We have that for every $t\geq 0$,
\begin{gather} \label{newqq}
G^{\varphi}_t=\QQphi(\vartheta > t\,|\,\cF_t)  = \frac{1}{\,^o(\eta^\varphi_t)}\cE_t(M^\FF\big)\cE_t(-\wtGaphi ).
\end{gather}
\el

\begin{proof}
Recall from the proof of Lemma \ref{lmbbb} that $Z^{\FF}= \cE(M^\FF)$ where $M^\FF$ is some $(\FF,\PP)$-martingale.
It will be useful to observe that
\begin{align*}
\,^o(\eta^\varphi)\cE(\wt N^{\varphi}) & = \cE(\wt N)\cE\big( \sTr{M^\FF} - G^{-1}_-\bcd \sTr{m}  \big) = \cE(G^{-1}_-\bcd m )\cE\big( \sTr{M^\FF} - G^{-1}_-\bcd \sTr{m}  \big) \\
& = \cE(G^{-1}_-\bcd m + \sTr{M^\FF} - G^{-1}_-\bcd \sTr{m}   + [G^{-1}_-\bcd m, \sTr{M^\FF} - G^{-1}_-\bcd \sTr{m}] \big) = \cE(M^\FF)
\end{align*}
where the penultimate equality follows from Yor's formula $\cE(X)\cE(Y)=\cE(X+Y+[X,Y])$ for arbitrary semimartingales $X$ and $Y$, and to obtain the last equality we have used the equalities $\wtG - G_-=\Delta m$ and $\Delta m\bcd [X, m]=[m,[X,m]]$ for any martingale $X$. Therefore, again by Lemma \ref{lem2.5}, the Az\'ema supermartingale of $\vt$ under $\QQphi$ is given by \eqref{newqq}.
\end{proof}

\brem
We remark that it is not clear whether $\vt$ is a {\it pseudo-stopping time} with respect to $\FF$ under $\QQ^\varphi$ since we were unable to either compute $\,^o(\eta^\varphi)$ or to verify whether the equality $\,^o(\eta^\varphi)=\cE(M^\FF\big)$ holds. For the definition and properties of pseudo-stopping times, we refer to Nikeghbali and Yor \cite{NY2005}. 
In particular, it is known from Theorem 1 of Aksamit and Li \cite{AL2016} that if a random time $\vartheta$ is an $\FF$-pseudo-stopping time under $\PP$, then $m=1$ in Lemma \ref{lem2.1} (i).
\erem

\section{Vulnerable European Options}     \label{sec3}

Our goal in this section is to give preliminary valuation results for the payoff occurring at a settlement date represented
by a $\GG$-stopping time $\whsigma$ but with a possible exogenous termination at a random time $\vt$, whichever comes first.
For brevity, the event of exogenous termination is referred to as the {\it default event}, which is consistent with the terminology used in models of the counterparty credit risk (for instance, for a vulnerable option or a defaultable swap) and credit derivatives, such as a credit default swap.

Observe that the processes introduced in Section \ref{sec2} depend on a choice of a probability measure.
In Section \ref{sec3}, all processes are invariably defined using a probability measure $\QQphi\in \cQ$ so that $G^{\varphi}_t:=\QQphi (\vt > t\,|\,\cF_t)$ and $\wtGphi_t:=\QQphi ( \vt \geq t\,|\,\cF_t)$. Similarly, the process $A^{o,\QQphi}$ (resp., $A^{p,\QQphi}$) is the dual $\FF$-optional (resp., $\FF$-predictable) projection under $\QQphi$ of the indicator process $A$.

We will first examine the structure of the vulnerable European option with the terminal payoff at time $\whsigma \wedge \vt$ when the payoff process $\whX$ is given by \eqref{eq3.1} and $\whsigma$ is an arbitrary stopping time from the class $\STG ={\cT}_{[0,T]}(\GG)$  of all $\GG$-stopping times with values in $[0,T]$.
Similarly, we denote by $\STF ={\cT}_{[0,T]}(\FF)$  the set of all $\FF$-stopping times with values in $[0,T]$.

\bd \label{def3.1}
The {\it payoff process} $\whX\in\cOgg$ is given by
\begin{equation} \label{eq3.1}
\whX:=P\I_{\llb 0,\vt\llb}+R_\vt\I_{\llb\vt,\infty\llb}
\end{equation}
where $P,R$ are bounded, $\FF$-optional processes. A {\it vulnerable European option} is a pair $(\whX_{\whsigma},\whsigma)$ where $\whsigma \in \STG$ and $\whX_{\whsigma}$ equals
\begin{equation} \label{eq3.2}
\whX_{\whsigma}:=P_{\whsigma}\I_{\llb 0,\vt\llb}(\whsigma)+R_{\vt}\I_{\llb \vt,\infty\llb}(\whsigma)
 = P_{\whsigma}\I_{\{\whsigma < \vt\}} +R_{\vt}\I_{\{\whsigma \geq \vt\}}.
\end{equation}
\ed

For any fixed $\sigma \in \STF$, the {\it stopped filtration} $\FF^\sigma$ is given by $\FF^\sigma:=(\cF_{\sigma \wedge t})_{t\geq 0}$.

\bl \label{lem3.1}
(i) For any $\whsigma \in \STG$, there exists $\sigma \in \STF$ such that $\whX_{\whsigma}=\whX_{\sigma\wedge\vt}=\whX_\sigma$. \\
(ii) For any $\sigma \in \STF $, there exists $\xx (\sigma)\in \mathcal{O}(\FF^\sigma)$ such that the equality $\whX_\sigma=\xx_\vt(\sigma)$ holds.
\el

\begin{proof}
(i) It is clear that the process $\whX$  given by \eqref{eq3.1} is stopped at $\vt$ so that $\whX=\whX^\vt$, which immediately implies that $\whX_{\whsigma}=\whX_{\whsigma \wedge \vt }$. Hence, by using also the well known property that, for any $\whsigma \in \STG$, there exist $\sigma \in \STF$ such that $\sigma \wedge \vt=\whsigma\wedge \vt$, we obtain the following equalities
\begin{align*}
\whX_{\whsigma}&=\whX_{\whsigma\wedge\vt}=P_{\whsigma\wedge\vt}\I_\seq{\whsigma\wedge\vt<\vt}+R_\vt\I_\seq{\whsigma\wedge\vt\geq\vt}= P_{\sigma\wedge\vt}\I_\seq{\sigma\wedge\vt<\vt}+R_\vt\I_\seq{\sigma\wedge \vt\geq\vt}\\
&=P_{\sigma}\I_\seq{\sigma<\vt}+R_\vt\I_\seq{\sigma\geq\vt}=\whX_{\sigma\wedge\vt}=\whX_\sigma .
\end{align*}
We conclude that $\whX_{\whsigma}=\whX_{\sigma\wedge\vt}=\whX_{\sigma}$ for some stopping time $\sigma \in \STF$, as was required to show.

\noindent (ii) It suffices to observe that
\begin{equation} \label{eq3.3}
\whX_{\sigma}=P_{\sigma}\I_{\rrb 0,\vt\llb}(\sigma)+ R_\vt\I_{\llb \vt,\infty \rrb}(\sigma)
=P_\sigma\I_{\rrb\sigma,\infty\llb}(\vt)+R_\vt\I_{\llb 0,\sigma\rrb}(\vt)=\xx_\vt(\sigma)
\end{equation}
where, for any fixed $\sigma\in\STF$, the $\FF$-adapted process $\xx(\sigma)$ is given by
\begin{equation} \label{eq3.4}
\xx(\sigma):=P_\sigma\I_{\rrb\sigma,\infty\llb}+R\,\I_{\llb 0, \sigma\rrb}=R^\sigma+(P_\sigma-R_\sigma)\I_{\rrb\sigma,\infty\llb}.
\end{equation}
Since the processes $P$ and $R$ are assumed to be $\FF$-optional, by Lemma 3.53 in He et al. \cite{HWY1992}, the process $\xx(\sigma)$ is $\FF^\sigma$-optional, although it is not a c\`adl\`ag process, in general. In view of equalities \eqref{eq3.3}, we will henceforth freely interchange the random variables $\whX_{\whsigma},\,\whX_{\sigma}$ and $\xx_\vt(\sigma)$
where $\sigma \in \STF$ is such that  $\sigma \wedge \vt=\whsigma\wedge \vt$.
\end{proof}

Let $\whsigma$ be a fixed, but otherwise arbitrary, stopping time from the class $\STG$ and let $\sigma \in \STF$ be a stopping time introduced in part (i) in Lemma \ref{lem3.1}. In view of the postulated boundedness of the payoff process $\whX$, the expected value appearing in Definition \ref{def3.2} below is well defined under any probability measure $\QQphi$ from $\cQ$.

\bd  \label{def3.2}
The {\it price process $\whV^{\QQphi}\in\cOgg$ under $\QQphi$} of the vulnerable European option $(\whX_{\whsigma},\whsigma)$ is the bounded $(\GG,\QQphi)$-martingale given by the equality $\whV^{\QQphi}_t:=\EQphi[\whX_{\whsigma}\,|\,\cG_t]=\EQphi \big[\whX_{\sigma \wedge \vt}\,|\,\cG_t\big]$ for every $t\in [0,T]$.
\ed

It is obvious that the process $\whV^{\QQphi}$ is stopped at the $\GG$-stopping time $\whsigma\wedge\vt=\sigma\wedge\vt$.
We wish to obtain a representation for the price process $\whV^{\QQphi}$ in terms of the processes $P,R$ and the Az\'ema supermartingale
of $\vt$ under $\QQphi$. The process $V^{\QQphi} := (\vv^{\QQphi})^{\sigma}$ where $\vv^{\QQphi}$ is the process introduced in Lemma \ref{lem3.2} is called the {\it $\FF$-reduced price} (or, simply, the {\it reduced price}) of $(\whX,\whsigma)$ under $\QQphi$.

\bl \label{lem3.2}
There exists a unique process $\vv^{\QQphi} \in\cOff$ such that $\whV^{\QQphi} \I_{\llb 0,\vt\llb}=\vv^{\QQphi} \I_{\llb 0,\vt \llb}$.
Moreover, $\vv^{\QQphi}_{\sigma}=P_{\sigma}$ and the following equality is valid
\begin{equation} \label{eq3.5}
\vv^{\QQphi} ={}^{o,\QQphi}\big(\whV^{\QQphi}\I_{\llb 0,\vt\llb}\big)(G^{\varphi})^{-1}.
\end{equation}
\el
\begin{proof}
The assertion follows easily from the theory of progressive enlargement of filtration (see, e.g., Proposition 2.8 and Lemma 2.9 in \cite{AJ2017}).
\end{proof}

As in Section \ref{sec2.0}, we formally identify any $\QQphi \in \cQ$ with the corresponding process $\varphi=(\varphi^{(o)},\varphi^{(pr)})$ and we recall that $\Phi$ is the set of all processes $(\varphi^{(o)},\varphi^{(pr)})$ associated with the class $\cQ$.
To alleviate notation, the process $\whV^{\QQphi}$ will be denoted as $\whV^{\varphi}$ and the same convention will be
applied to processes $V^{\QQphi}$ and $\vv^{\QQphi}$ so that we will write $V^{\varphi}$ and $\vv^{\varphi}$, respectively.

\bp \label{pro3.1}
The price process $\whV^{\varphi}$ satisfies
\begin{equation*}
\whV^{\varphi}=\I_{\llb 0,\vt\llb}\bcd V^{\varphi}+(R_{\vt}-V^{\varphi}_{\vt-})\I_{\llb \vt,\infty\llb}\I_\seq{\sigma\geq\vt}.
\end{equation*}
\ep

\begin{proof}
Let $\QQphi \in \cQ$ be fixed and let us consider the $(\QQphi,\FF)$-martingale $\Mxt$ defined by $\Mxt_t:=\EE_{\QQphi}[(\xx (\sigma)\bcd A^{o,\QQphi})_\infty|\cF_t]$
where the process  $\xx(\sigma)\in \mathcal{O}(\FF^\sigma)$ is given by \eqref{eq3.4}. Then the following equalities hold on $\llb 0,\sigma\rrb$
\begin{equation}  \label{eq3.6}
\Mxt={}^{o,\QQphi}\big(P_{\sigma} G^{\varphi}_{\sigma}+(R\bcd A^{o,\QQphi})_{\sigma}\big).
\end{equation}
Furthermore, the equality
\begin{equation}  \label{eq3.7}
{}^{o,\QQphi}\big(\whV^{\varphi} \I_{\llb 0,\vt\llb}\big)=\Mxt-R\bcd A^{o,\QQphi}
\end{equation}
holds on $\llb 0,\sigma\rrb$. Using equations \eqref{eq3.5} and \eqref{eq3.7}, we represent the process $\vv^{\varphi}$ on $\llb 0, \sigma \rrb$ as follows
\begin{equation} \label{eq3.9}
\vv^{\varphi}=(G^{\varphi})^{-1}(\Mxt-R\bcd A^{o,\QQphi})
\end{equation}
and we may now assume, without loss of generality, that the process $\vv^{\varphi}$ is stopped at $\sigma $. If we set $V^{\varphi}:=(\vv^{\varphi})^\sigma$,
then it is obvious that $\whV^{\varphi}\I_{\llb 0,\sigma\rrb\cap\llb 0,\vt\llb}=V^{\varphi}\I_{\llb 0,\sigma\rrb\cap\llb 0,\vt\llb}$
and $V^{\varphi}_{\sigma}=P_{\sigma}$. Using the equality
\[
\xx_{\vt}(\sigma)\I_{\llb\vt,\infty\llb\cup\rrb\sigma,\infty\llb}
=P_{\sigma}\I_{\rrb\sigma,\infty\llb}\I_\seq{\sigma<\vt}+R_{\vt}\I_{\llb \vt ,\infty \llb}\I_\seq{\sigma\geq\vt},
\]
we obtain
\begin{align*}
\whV^{\varphi}&=V^{\varphi}\I_{\llb 0,\sigma \rrb \cap \llb 0,\vt\llb}+\xx_{\vt}(\sigma)\I_{\llb\vt,\infty \llb \cup \rrb\sigma,\infty\llb}
=\I_{\llb 0,\vt \llb}\bcd V^{\varphi}+(\xx_{\vt}(\sigma)-V^{\varphi}_{\vt-})\I_{\llb\vt,\infty \llb \cup \rrb\sigma,\infty\llb}\\
&=\I_{\llb 0,\vt \llb}\bcd V^{\varphi}+(R_{\vt}-V^{\varphi}_{\vt-})\I_{\llb\vt,\infty \llb}\I_\seq{\sigma \geq \vt}
\end{align*}
since $V^{\varphi}_{\vt-}=V^{\varphi}_{\sigma}=P_{\sigma}=\xx_{\vt}(\sigma)$ on the event $\seq{\sigma<\vt}$ and, by the usual convention, $V^{\varphi}_{0-}=0$.
\end{proof}

\subsection{Upper Price of a Vulnerable European Option}           \label{sec3.1.2}

We adopt the following definition of the {\it upper price} of a vulnerable European option $(\whX_{\whsigma},\whsigma)$. Notice that we also use here Lemma \ref{lem3.1} and we observe that since the payoff process is bounded the upper price $\whVu$ is well defined.
The main result on the upper price  of a vulnerable European option is Theorem \ref{th3.1} where it is shown that the $\FF$-reduced
upper value $V^u$ coincides with the value process of a constrained optimal stopping problem under the unique ELMM $\QQ$ for the market model $(S,\FF,\PP)$.

\bd \label{def5.1}
The {\it upper price} of a vulnerable European option $(\whX_{\whsigma},\whsigma)$ equals, for every $t\in [0,T]$,
\begin{equation*}
\whVu_t:=\esssup_{\varphi\in\Phi}\whV^{\varphi}_t=
\esssup_{\QQphi\in\cQ}\EQphi[\whX_{\whsigma}\,|\,\cG_t]=\esssup_{\QQphi\in\cQ}\EQphi \big[\whX_{\sigma\wedge\vt}\,|\,\cG_t\big]
=\esssup_{\QQphi\in\cQ}\EQphi \big[\xx_{\vt}(\sigma)\,|\,\cG_t\big]
\end{equation*}
and the {\it $\FF$-reduced upper price} $\VVu$ is given by the equality $\VVu=(\vv^u)^{\sigma}$ where $\vv^u$ is a unique $\FF$-optional process
such that $\whVu\I_{\llb 0,\vt\llb}=\vv^u\I_{\llb 0,\vt\llb}$ where $\sigma \in \STF$ is such that $\whsigma\wedge\vt=\sigma\wedge\vt$.
\ed

We deduce from \eqref{eq3.5} that $\vv^u$ has the following representation
\begin{equation} \label{eq5.1}
\vv^u_t=\esssup_{\QQphi\in\cQ}\frac{\EQphi \big[\whX_{\sigma\wedge\vt}\I_{\{\vt>t\}}\,|\,\cF_t\big]}{G^\varphi_t}
 =\esssup_{\QQphi\in\cQ} \vv^{\QQphi}_t = \esssup_{\varphi \in\Phi} \vv^{\varphi}_t .
\end{equation}

Our next goal is to obtain a more explicit representation for the right-hand side in \eqref{eq5.1} through the class of Radon-Nikodym densities with respect to $\PP$ of probability measures from $\cQ$. To this end, we introduce the following notation:
\begin{align*}
H^{\varphi}& :=\,^o(\eta^\varphi)G^{\varphi}, \qquad \wtH^{\varphi}:=\,^o(\eta^\varphi)\wtGphi, \qquad \wt D^{\varphi} := \wtH^\varphi \bcd \wtGamma^\varphi
\end{align*}
where the process $\wtGaphi= \big(1+ \varphi^{(o)}(1-\Delta \wtGamma)\big) \bcd \wtGamma$ is the $\FF$-optional hazard process of $\vt$ under $\QQphi$ (see Proposition \ref{pro2.2}).

\bl \label{lem5.1}
The process $\vv^u$ given by \eqref{eq5.1} satisfies $\vv^u_t=\esssup_{\,\varphi \in\Phi}\vv^{\varphi}_t$ where, on the event $\{\sigma\geq t\}$,
\begin{equation}
\vv^{\varphi}_t=(G^{\varphi}_t)^{-1}\,\EQphi \Big[G^{\varphi}_{\sigma}P_{\sigma}+\int_{\rrb t,\sigma \rrb} R_s\,dA^{o,\QQphi}_s\,\Big|\,\cF_t\Big] \label{eq5.2.1}
\end{equation}
or, equivalently,
\begin{align}  \label{eq5.2}
\vv^{\varphi}_t=(H^{\varphi}_t)^{-1}\,\EP\Big[H^{\varphi}_{\sigma}P_{\sigma}+\int_{\rrb t,\sigma \rrb}R_s \,d\wt D^{\varphi}_s\,\Big|\,\cF_t\Big].
\end{align}
\el

\begin{proof}
In view of equation \eqref{eq5.1}, the first asserted equality follows from computations analogous to those in the proof of Proposition \ref{pro3.1}. To deduce equation \eqref{eq5.2} from the first equality, it suffices to use results from Section \ref{sec2.2}, the abstract Bayes formula for a fixed $\QQphi \in \cQ$, equation \eqref{eq2.14} and Proposition \ref{pro2.2}.
\end{proof}

\subsection{Reduced Upper Price via Constrained Optimal Stopping Problem}   \label{sec5.3a}

If almost all sample paths of a continuous time process $B$ are nondecreasing functions on $[0,T]$, then we denote by $\xS^r(B)$ the {\it right support} of $B$, which is defined by, for almost every $\omega$,
\begin{align*}
\xS^r(B):= \big\{ t\in \RR_+ : \forall\,\varepsilon>0, \ B_{t+\varepsilon}-B_{t}>0  \big\}
\end{align*}
where, by convention, $B_t=0$ for $t<0$ and $B_t=B_T$ for $t>T$.
 In particular, given a random time $\vt$, we recall that $A^o$ is the dual $\FF$-optional projection of $A = \I_{\llb \vt , \infty \llb}$.
We assume that $A^o$ is continuous and we denote by $\xS^r(A^o)$ the right support of the increasing process $A^o$.
Finally, we define $\barS(A^o):=\xS^r(A^o)\cup \{T\}$ and we will frequently write $\barS$ instead of $\barS(A^o)$.
It is clear that under Assumption \ref{ass2.1} (i) we have that $\barS (A^o) = \barS (\wtGamma )$.

We need to introduce additional notation for some classes of stopping times. Let $\bSTF$ (resp., $\bSTFp$) stand for the class of all $\FF$-stopping times  (resp., $\FF$-predictable stopping times) $\tau$ taking values in $[0,T]$ and such that $\PP ( \tau \in \barS)=1$. Similarly, $\bSTFt$ is the set of $\FF$-stopping times from $\bSTF$ with values in the set $\barS \cap [t,T]$.

\bd
We denote by $\barV^{\FF}$ be the value process of the following constrained optimal stopping problem under
$\QQ$, for every $t\in [0,T]$
\begin{align} \label{osp1}
\barV^{\FF}_t=\esssup_{\tau\in \bSTFt}\EQ\big[P_{T}\I_{\{\tau=T\}}+R_{\tau}\I_{\{\tau< T\}}\,|\,\cF_t\big].
\end{align}
\ed

To analyze the upper price of a vulnerable European option via the constrained optimal stopping problem \eqref{osp1}, we need to show that the process
$\barV^{\FF}$ is related to a sequence of penalized BSDEs \eqref{eq1.29}. To this end, we will make use of results from Li et al. \cite{LLR2} on generalized BSDEs (GBSDEs).


We introduce the space $\cH^2(\whMM)=\{ Z \in \cPff : \|Z\|^2_{\cH^2(\whMM)}< \infty \}$ with the pseudo-norm
\begin{align*}
\|Z\|_{\cH^2(\whMM)}:=\Big[ \EQ\big[(Z^2\bigcdot [\whMM])_T\big] \Big]^{1/2}
\end{align*}
where the $\FF$-martingale $\whMM$ given in equation \eqref{MQ} is assumed to be square-integrable so that $\EP [[\whMM]_T] < \infty$.
It is known that the space $\cS^2(\FF)=\{ X \in \cOff : \|X\|^2_{\cS^2(\FF)}< \infty \}$ with the norm
\begin{align*}
\|X\|_{\cS^2(\FF)}:=\bigg[ \EQ\bigg(\esssup_{\tau \in \STF}|X_\tau|^2\bigg)\bigg]^{1/2}
\end{align*}
is a Banach space (see Proposition 2.1 in Grigorova et al. \cite{GIOOQ2017}).
If $X$ and $Y$ are arbitrary $\FF$-optional processes, then the inequality $Y\geq X$ means that $Y_{\sigma}\geq X_{\sigma}$ for every $\sigma \in \STF$.

\bhyp \label{ass3.1}
We postulate that: \\
(i) the processes $P$ and $R$ are  $\FF$-optional, nonnegative and bounded; \\
(ii) the process $\wtGamma$ (and hence also the process $\wtGamma^\varphi$ for every $\varphi \in \Phi$) is continuous and bounded.
\ehyp

Alternatively, Assumption \ref{ass3.1} (ii) can be replaced by the following weaker condition:\\
(ii') the process $\wtGamma$ is continuous and satisfies $\EE[e^{\wtGamma_{T\wedge \vartheta}}] < \infty$ and $\EQ[\wtGamma_T^2] < \infty$.

The postulate that the optional hazard process $\wtGamma$ is continuous (i.e., $\wtGamma = \wtGamma^c$) is in our opinion not essential and is only made for ease of presentation. In fact, we conjecture that it is possible to obtain results similar to Theorems \ref{th3.1}, \ref{th6.1} and \ref{th6.2} in the case where $\wtGamma$ is purely discontinuous (i.e., when $\wtGamma= \wtGamma^d$) and we foresee that the mixed case is also not out of reach. Without entering into the details, as it is still part of ongoing works, we anticipate that similar to Theorem \ref{th3.1}, if the $\FF$-optional hazard process $\wtGamma$ is purely discontinuous, then the upper price of a vulnerable European option is equal to the price of an option of Bermudan style with the exercise dates coinciding with jump times of the optional hazard process. 


%

Recall from Section \ref{sec2.0} that $\QQ$ is a unique ELMM for the market model $(S,\FF,\PP)$. It appears that the process  $V^{\varphi} = (v^\varphi)^\sigma$ has a simple representation through a generalized BSDE under $\QQ$ as shown by the following result.

\bp \label{pro5.1x}
For any fixed $\varphi \in \Phi$, there exists an $\FF$-predictable process $Z^\varphi$ such that $(V^{\varphi}, Z^\varphi)$ belongs to $\cS^2(\FF)\times \cH^2(M^\QQ)$ and satisfies, on the interval $[0,\sigma]$, the generalized BSDE
\begin{equation} \label{eq5.6a}
V^{\varphi}_t=P_{\sigma}+\int_{\rrb t,\sigma \rrb}(R_s-V^{\varphi}_s)\,d\wtGamma^\varphi_s
-\int_{\rrb t,\sigma \rrb} Z^\varphi_{s}\,d\whMM_s.
\end{equation}
\ep

\begin{proof}
In view of \eqref{newqq} the process  $H^\varphi=\,^o(\eta^\varphi)G^{\varphi}$ satisfies the equality $H^\varphi = \cE(M^\FF)\cE(-\wtGaphi)$ and thus, using  the  properties of the dual $\FF$-optional projection and the martingale property of $Z^\FF = \cE(M^\FF)$, we obtain from \eqref{eq5.2}
\begin{align}
\vv^{\varphi}_t
				& =\dfrac{\EP\Big[\cE_\sigma(M^\FF)\cE_\sigma(-\wtGamma^{\varphi}) P_{\sigma}+\int_{\rrb t,\sigma \rrb}R_s \,\cE_s(M^\FF\big)\cE_s(-\wtGamma^{\varphi}_-) \,d\wtGamma^\varphi_s \,\Big|\,\cF_t\Big]}{\cE_t(M^\FF)\cE_t(-\wtGamma^{\varphi}\big)}\nonumber\\
				& =\dfrac{\EP\Big[\cE_\sigma(M^\FF\big)\left[\cE_\sigma(-\wtGamma^{\varphi}\big) P_{\sigma}+\int_{\rrb t,\sigma \rrb}R_s \cE_s(-\wtGamma^{\varphi}_-) \,d\wtGamma^\varphi_s\right] \,\Big|\,\cF_t\Big]}{\cE_t(M^\FF\big)\cE_t(-\wtGamma^{\varphi}\big)}\nonumber \\
				& =\dfrac{\EQ\Big[\cE_\sigma(-\wtGamma^{\varphi}\big) P_{\sigma}+\int_{\rrb t,\sigma \rrb}R_s \cE_s(-\wtGamma^{\varphi}_-) \,d\wtGamma^\varphi_s \,\Big|\,\cF_t\Big]}{\cE_t(-\wtGamma^{\varphi}\big)}. \label{eq3.18}
\end{align}

If we denote $\zeta^{\varphi}:= \cE(-\wtGamma^{\varphi}\big)$, then the last equality gives
\begin{align*}
\vv^{\varphi}_t =\big(\zeta^{\varphi}_t\big)^{-1}\,  \EQ\Big[\zeta^{\varphi}_{\sigma} P_{\sigma}+\int_{\rrb t,\sigma \rrb}R_s
\zeta^{\varphi}_{s-} \,d\wtGamma^\varphi_s \,\Big|\,\cF_t\Big].
\end{align*}
Let $X:=M^{\varphi}-R \zeta_-^{\varphi} \bcd \wtGamma^{\varphi}$ where
\begin{align*}
M^{\varphi}_t := \EQ\Big[\zeta^{\varphi}_{\sigma} P_{\sigma}+\int_{\rrb 0,\sigma \rrb}R_s
\zeta^{\varphi}_{s-} \,d\wtGamma^\varphi_s \,\Big|\,\cF_t\Big]
\end{align*}
so that $\vv^{\varphi} =\big(\zeta^{\varphi}\big)^{-1}X$. Then the It\^o integration by parts formula gives
\begin{align*}
\vv^{\varphi}&= \vv^{\varphi}_0 + (\zeta_-^{\varphi})^{-1}\bcd X
+ X_-\bcd (\zeta^{\varphi})^{-1} + [X,(\zeta^{\varphi})^{-1}] \\
&=\vv^{\varphi}_0+ (\zeta_-^{\varphi})^{-1} \bcd X + X \bcd (\zeta^{\varphi})^{-1}  = \vv^{\varphi}_0+ (\zeta_-^{\varphi})^{-1}\bcd X + X (\zeta^{\varphi})^{-1}\bcd \wtGamma^\varphi \\
&=\vv^{\varphi}_0+ (\zeta_-^{\varphi})^{-1} \bcd M^{\varphi} - R \bcd\wtGamma^{\varphi}  + \vv^{\varphi} \bcd \wtGamma^\varphi =
\vv^{\varphi}_0 + (\zeta_-^{\varphi})^{-1} \bcd M^{\varphi} - ( R -\vv^{\varphi} )\bcd \wtGamma^\varphi .
\end{align*}
Next we show that $(\zeta_-^{\varphi})^{-1} \bcd M^{\varphi}$ is uniformly integrable martingale under $\QQ$.
Using the fact that $R$ and $P$ are bounded, we can easily deduce that $v^\varphi$ is bounded for any $\varphi \in \Phi$. To see that
\begin{gather*}
\vv^{\varphi}_t \leq c\EQ\Big[e^{-(\wtGamma^\varphi_\sigma -\wtGamma^\varphi_t)} +e^{\wtGamma^\varphi_t}  \int_{\rrb t,\sigma \rrb}
e^{-\wtGamma^\varphi_s} \,d\wtGamma^\varphi_s \,\Big|\,\cF_t\Big] \leq 2c.
\end{gather*}
then there exists some positive constant $C$ such that
\begin{gather*}
|(\zeta_-^{\varphi})^{-1} \bcd M^{\varphi}| \leq  |v^\varphi - \vv^{\varphi}_0 + (R-v^\varphi)\bcd \wtGamma^\varphi| \leq C\wtGamma
\end{gather*}
since $R$, $\vv^\varphi$ and $\varphi^{(o)}$ are bounded. This shows that $(\zeta_-^{\varphi})^{-1} \bcd M^{\varphi}$ is square-integrable 
under $\QQ$  as long as $\wtGamma$ is $\QQ$ square-integrable.
We recall (see Assumption \ref{ass2.1} (iii)) that the $\FF$-martingale $\whMM$ have the predictable representation property under $\QQ$ and
thus $M^{\varphi} = \psi^{\varphi} \bcd \whMM$ for some $\FF$-predictable process $\psi^{\varphi}$, which completes the derivation of \eqref{eq5.6a} since $V^{\varphi}_{\sigma} =P_{\sigma}$.
\end{proof}

\brem 
The linear generalized BSDE \eqref{eq5.6a} can also be formally derived by applying Lemma 7.3 in \cite{ALR2022} with $G = \cE(-\wtGamma),\, \wtG = \cE(-\wtGamma_-),\, m = 1$ and $A^o = 1-\cE(-\wtGamma)$. It is then useful to stress that it is not clear whether the equality $m = 1$ holds in the present setup, although algebraically the computations in the proof of Proposition \ref{pro5.1x} are analogous to those from the proof
of Lemma 7.3 in \cite{ALR2022}.
\erem

The following corollary to Propositions  \ref{pro5.1x} gives an explicit form of the unique solution
to the GBSDE \eqref{eq5.6a} when the $\FF$-optional hazard process $\wtGamma$ is continuous.

\bcor \label{cor5.9}
Let $(V^\varphi, Z^\varphi) \in \cS^2(\FF) \times \cH^2(M^\QQ)$ be a solution on the interval $[0,\sigma]$ to the GBSDE
\begin{align}  \label{eq5.36}
V^{\varphi}_t=P_\sigma+\int_{\rrb t,\sigma\rrb} \big(\varphi^{(o)}_s +1\big)\big(R_s-V^{\varphi}_s\big)\,d\wtGamma_s-\int_{\rrb t,\sigma\rrb}Z^{\varphi}_s\,d\whMM_s.
\end{align}
Then $V^\varphi = (v^\varphi)^\sigma$ and
\begin{align}  \label{eq5.35}
(\vv^\varphi_t)^\sigma = V^{\varphi}_t=\EQ\big[P_\sigma\cE_{t,\sigma}(-\wtGamma^\varphi)+(\I_{\rrb t,\sigma\rrb }R_t
\cE_{t,\cdot}\big(-\wtGamma^\varphi)\bcd \wtGamma^{\varphi}\big)_\sigma\,|\,\cF_t\big].
\end{align}
\ecor

\begin{proof}
As $\wtGamma$ is continuous, it suffices to apply integration by parts to $\cE(\wtGamma) V^{\varphi}$.
\end{proof}

Notice that the process $\varphi^{(o)}\in \cOff$ can be taken to
be any bounded process such that $\varphi^{(o)} +1 > 0$, for any $n \in \NN$, we may take $\varphi^{(o)}=n-1=:\varphi^n$ such that $\varphi=\{\varphi^n, \varphi^{(pr)}\} \in \Phi^n$  where $\Phi^n$ is a non-empty subset of $\Phi$. In the following, for simplicity of presentation, we henceforth assume that $\whsigma=\sigma=T$ but that assumption is not essential when dealing with vulnerable European options.
The next result furnishes a link between a constrained optimal stopping problem and a sequence of penalized GBSDEs.

\bp \label{npro1.15}
Let $R$ be a right-continuous process. For every $n\in\NN$, let $(\wtY^n,\wtZ^n)$ be the unique solution in $\cS^2(\FF)\times \cH^2(\whMM)$ to the following GBSDE under $\QQ$ on $[0,T]$
\begin{align} \label{eq1.29}
\wtY^n_t=P_T+\int_{]t,T]} n(R_s-\wtY^n_s)^+\,d\wtGamma_s-\int_{]t,T]}\wtZ^n_s\,d\whMM_s .
\end{align}
Then the sequence $(\wtY^n)_{n\in \NN}$ of stochastic processes is increasing and converges almost surely to the value process $\barV^{\FF}$ of the constrained
optimal stopping problem \eqref{osp1}.
\ep

\begin{proof}
For a fixed $n \in \NN$, we note that the existence and uniqueness of a solution $(\wtY^n, \wtZ^n)\in \cS^2(\FF)\times \cH^2(\whMM)$ to the GBSDE \eqref{eq1.29} follows from Proposition 3.5 in \cite{LLR2}. Additionally, from the comparison theorem for GBSDE given by Proposition 3.1 in \cite{LLR2}, we have $\wtY^n \leq \wtY^{n+1}$ for $n\in \NN$.

In the first step, we will show that the c\`adl\`ag, $\FF$-adapted process $\wtY^n$  satisfies, for every $n \in \NN$ and all $t\in [0,T]$,
\begin{align} \label{eq1.30}
\wtY_t^n=\esssup_{\tau\in \bSTFt}\EQ\big[P_{T}\I_{\{\tau=T\}}+(R_{\tau}\land\wtY_{\tau}^n)
\I_{\{\tau< T\}}\,|\,\cF_t\big].
\end{align}
For a fixed $n\in \NN$, we define $\yxi^n_t :=  P_{T}\I_{\{t=T\}}+(R_{t}\land\wtY_{t}^n)\I_{\{ t < T\}}$ and we observe that
$$\yxi^n_t = (P_{T}\wedge \wtY_T^n)\I_{\{t=T\}}+(R_{t}\land\wtY_{t}^n)\I_{\{ t < T\}} \leq \wtY_t^n.$$
Furthermore, the GBSDE \eqref{eq1.29} can be represented as
\begin{align} \label{eq1.31}
\wtY^n_t=P_T-\int_{]t,T]}\wtZ^n_s\,d\whMM_s+\wtK^n_T-\wtK^n_t
\end{align}
where the $\FF$-adapted, continuous, nondecreasing process $\wtK^n$ is given by
\begin{align*}
\wtK^n_t:=\int_{]0,t]} n(R_s-\wtY^n_{s})^+\,d\wtGamma_s .
\end{align*}
Recall that we assumed that $R$ (and hence also $\yxi^n$) is a right-continuous process. We claim that \eqref{eq1.30} is valid,
that is, for every $t\in [0,T]$,
\begin{align} \label{eq1.32}
\wtY_t^n=\esssup_{\tau\in \bSTFt}\EQ\big[\yxi^n_{\tau}\,|\,\cF_t\big].
\end{align}
Equality \eqref{eq1.32} is obvious for $t=T$ so it suffices to consider any $t<T$. We have, for any $\tau \in \bSTFt$,
\begin{align*}
\wtY_t^n =\EQ\big[ \wtY^n_{\tau}+\wtK^n_{\tau}-\wtK^n_t  \,|\,\cF_t\big]
\geq \EQ\big[ \wtY^n_{\tau} \,|\,\cF_t\big] \geq \EQ\big[ \yxi^n_{\tau} \,|\,\cF_t\big],
\end{align*}
which implies that $\wtY_t^n \geq \esssup_{\tau\in \bSTFt}\EQ\big[\yxi^n_{\tau}\,|\,\cF_t\big]$.

For the converse inequality, we define the stopping time $\tau_t:=\inf\{s\in [t,T]\,|\, \wtK^n_s-\wtK^n_t>0\}\in \bSTFt$
and we observe that, due to the continuity of $\wtK^n$, we have that $\wtK^n_{\tau_t}-\wtK^n_t=0$.
Furthermore, on the event $\{\tau_t< T\}$ we have $\wtK^n_s>\wtK^n_{\tau_t}$ on $\rrb \tau_t,T\rrb$, which
entails the inequality $\limsup_{u \downarrow \tau_t}(R_u-\wtY^n_u)\geq 0$ and thus, since $R$
is right-continuous and $\wtY^n$ is right-continuous, we conclude that $R_{\tau_t}-\wtY^n_{\tau_t}\geq 0$,
which in turn implies that $\wtY^n_{\tau_t}= \yxi^n_{\tau_t}$.
On the event $\{\tau_t = T\}$ we have that $\wtK^n_{\tau_t}-\wtK^n_t=\wtK^n_T-\wtK^n_t=0$ and
$\wtY^n_{\tau_t}=\yxi^n_{\tau_t}$. From \eqref{eq1.31}, we now get
\begin{align*}
\wtY^n_t=\wtY^n_{\tau_t}-\int_{\rrb t,\tau_t\rrb}\wtZ^n_s\,d\whMM_s=\yxi^n_{\tau_t}-\int_{\rrb t,\tau_t\rrb}\wtZ^n_s\,dM_s
= \EQ\big[ \yxi^n_{\tau_t}\,|\,\cF_t\big],
\end{align*}
which leads to the inequality $\wtY_t^n\leq \esssup_{\tau\in \bSTFt}\EQ\big[\yxi^n_{\tau}\,|\,\cF_t\big]$ since $\tau_t \in \bSTFt$.
We conclude that equality \eqref{eq1.32} holds.

In the second step, we set $\xi_t:=P_{T}\I_{\{t=T\}}+R_{t}\I_{\{t< T\}}$ and we show that $\lim_{\,n\rightarrow\infty} \wtY_t^n = \barV^{\FF}_t$ where
\begin{align}  \label{eq1.34}
\barV^{\FF}_t :=\esssup_{\nu \in \bSTFt}  \EQ\big[\xi_{\nu}\,|\,\cF_t\big].
\end{align}
It follows from  \eqref{eq1.30} and \eqref{eq1.34} that $\wtY^n \leq \barV^{\FF}$ and  $\wtY^n \uparrow \barY \leq \barV^{\FF}$ where the $\FF$-optional process $\barY$ is given by $\barY=\lim_{\,n\rightarrow +\infty}\wtY^n$. Furthermore, it is clear that the
process $\barY$ is nonnegative and belongs to $\cS^2(\FF)$ since it is dominated by $\barV^{\FF}\in \cS^2(\FF)$. Furthermore, to show that $\barY$ dominates $\xi$ on $\barS$ we note that, for every $\tau \in \bSTF$,
\begin{align*}
\wtY_\tau^n& \geq  \barY^n_\tau = P_T+ \int_{\rrb \tau, T\rrb} n(R_s- \barY^n_s)\,d\wtGamma_s-\int_{\rrb\tau,T\rrb} \barZ^n_s\,d\whMM_s
\end{align*}
where the right-hand side is a linear BSDE and we have, for every $\tau \in \bSTFt$,
\begin{align*}
\barY_\tau^n= \EQ\big[P_{T}\cE_{\tau,T}(-\wtGamma^n)+(\I_{\rrb \tau,T\rrb}R\cE_{\tau,\cdot}(-\wtGamma^n)\bcd \wtGamma^n)_{T}\,|\,\cF_\tau\big]
\end{align*}
where $\wtGamma^n= n \wtGamma$. We observe that $\cE_{\tau,T}(-\wtGamma^n)$ converges to $\I_{\{\tau=T\}}$ as $n\rightarrow \infty$ and thus, using Lemma \ref{lem3.4x} and the right-continuity of $R$, we obtain
\begin{align}
\barY_\tau=\lim_{\,n\rightarrow \infty}\wtY^n_{\tau} \geq \lim_{\,n \rightarrow \infty} \barY^n_{\tau}=P_{\tau}\I_{\{\tau= \sigma\}}+R_{\tau}\I_{\{\tau< \sigma\}} = \xi_\tau. \label{VR}
\end{align}
Thus $\barY$ dominates $\xi$ on $\barS$. To conclude that $\barY = \barV^{\FF}$, it now suffices to show that  $\barY \geq \barV^{\FF}$.

To this end, by using the fact that the essential supremum and conditional expectation can be interchanged, we can conclude that $\barV^{\FF}$ is the smallest strong supermartingale dominating $\xi$ on $\barS$. From the monotone convergence theorem we obtain, for every $0\leq \tau \leq \nu \leq T$,
\begin{gather*}
 \EQ\big[ \barY_\nu \,|\, \cF_\tau \big] = \lim_{\,n\rightarrow \infty} \EQ\big[\wtY^n_\nu \,|\,\cF_\tau \big]
 \leq \lim_{\,n\rightarrow \infty} \wtY^n_\tau = \barY_\tau.
\end{gather*}
Therefore, $\barY$ is a strong $\FF$-supermartingale dominating $\xi$ on $\barS$, which in turn implies that,
for every $t\in[0,T]$,
\begin{gather*}
\barY_t \geq \esssup_{\nu \in \bSTFt} \EQ\big[\xi_{\nu}\,|\,\cF_t\big]=\barV^{\FF}_t ,
\end{gather*}
as was required to show.
\end{proof}

The following auxiliary result is an extension of Lemma 6.1 in \cite{EKPPQ1997}.

\bl  \label{lem3.4x}
Suppose that $R$ is a bounded and right-continuous process. Then for any $\FF$-stopping time $\nu$ such that
$\PP( \nu \in \barS\cap[0,T])=1$, we have
\begin{align*} 
\lim_{n\to\infty}P_T\cE_{\nu,T}(-\wtGamma^n)+(\I_{\rrb \nu,T\rrb}R\cE_{\nu,\cdot}(-\wtGamma^n)\bcd \wtGamma^n)_T\
=P_{\nu}\I_{\{\nu=T\}}+R_{\nu}\I_{\{\nu<T\}}
\end{align*}
where $\wtGamma^n = n\wtGamma$.
\el

\begin{proof}
By assumption, $\wtGamma$ is continuous and $R$ is a bounded and right-continuous process. Let us fix a stopping time $\nu$ such that $\PP( \nu \in \barS\cap[0,T])=1$. For any $n\in\NN$, we recall that $\varphi^{n} =n-1 $ and
\begin{align*}
\wtGamma^n:=\wtGamma^{\varphi^{n}}=(1+\varphi^{n})\bcd\wtGamma =n \wtGamma .
\end{align*}
Since $\wtGamma^n_\nu-\wtGamma^n_T=0$ we have, on the event $\{\nu=T\}$,
\begin{align*}
\lim_{n\to\infty} \cE_{\nu,T}(-\wtGamma^n)=\lim_{n\to\infty} e^{\wtGamma^n_\nu-\wtGamma^n_T}=\I_{\{\nu=T\}}.
\end{align*}

Let us now consider the event $\{\nu<T\}$. Since $\nu$ has values in the right support of $\wtGamma$ we deduce that $\wtGamma_T - \wt\Gamma_\nu > 0$ and hence $\lim_{n\to\infty} (\wtGamma^n_\nu-\wtGamma^n_T)=-\infty$. We claim that, on the event $\{\nu<T\}$,
\begin{align}
\lim_{n\to\infty}\int_{\rrb\nu,T\rrb}R_s e^{\wtGamma^n_{\nu}-\wtGamma^n_s}\,d\wtGamma^n_s
=R_{\nu}\I_{\{\nu<T\}}\label{dirac}
\end{align}
since the sequence of measures $\I_{\rrb\nu,T\rrb} e^{\wtGamma^n_{\nu}-\wtGamma^n_s}\,d\wtGamma^n_s$ converges to the Dirac measure at
$\nu$ on the event $\{\nu<T\}$, that is, to the measure $\I_{\{\nu<T\}}\delta_{\nu}$.
The convergence in \eqref{dirac} can be checked by using the time change on $[0,T]$ associated with $\wtGamma$. If we define the increasing, right-continuous process $C$ by $C_s=\inf\{t\in\RR_+:\wtGamma_t>s\}$, then an application of the time change formula (see, e.g., Chapter 0 in Revuz and Yor \cite{RY1999}) gives
\begin{align*}
\int_{\rrb\nu,T\rrb}R_se^{\wtGamma^n_{\nu}-\wtGamma^n_s}\,d\wtGamma^n_s
&=\int_{\rrb\nu,T\rrb}R_s ne^{n(\wtGamma_{\nu}-\wtGamma_s)}\,d\wtGamma_s\\
&=\int_0^\infty \I_{\{\nu< C_s \leq T\}}  R_{C_s} n e^{n(\wtGamma_{\nu}-s)}\,ds
\end{align*}
where in the second equality, we have used the fact that $\wtGamma_{C_s} =s$ since $\wtGamma$ is continuous. From the fact that $\{\nu < C_s\} \subseteq \{\wtGamma_\nu \leq s\}$ and the change of variable $u = s- \wtGamma_\nu$ we obtain
\begin{align*}
\int^\infty_{0} \I_{\{\nu< C_s \leq T\}}  R_{C_s} n e^{n(\wtGamma_{\nu}-s)}\,ds &=\int^\infty_0 \I_{\{\nu< C_{\wtGamma_\nu+ u} \leq T\}}  R_{C_{\wtGamma_\nu+ u}} n e^{-nu}\,du\\
&=\EE_X\big[\I_{\{\nu< C_{\wtGamma_{\nu+ X/n}}\leq T\}}R_{C_{\wtGamma_{\nu+ X/n}}}\big]
\end{align*}
where in the last equality we have use the fact that $ne^{-nu}$ is the density of $\frac{1}{n}X$ where has a unit exponential distribution
and is independent of the $\sigma$-field $\cG_\infty$.

Recalling that $\nu< T$ has values in the right support of $\wtGamma$, we deduce that $\wtGamma_{\nu + X/n} \geq \wtGamma_{\nu + X/(n+1)}  > \wtGamma_{\nu}$ for all $n \in \mathbb{N}$. This observation, together with the right-continuity of processes $R,C$ and $\wtGamma$, gives
\begin{align*}
\lim_{n\rightarrow \infty} \EE_X\big[\I_{\{\nu< C_{\wtGamma_\nu+ X/n} \leq T\}}R_{C_{\wtGamma_\nu+ X/n}}\big] = \I_{\{\nu \leq C_{\wtGamma_\nu} < T\}}  R_{C_{\wtGamma_\nu}}.
\end{align*}
Finally, since $\nu < T$ and $\nu$ takes values in the right support of $\wtGamma$ we have $C_{\wtGamma_\nu} = \inf\{s: \wtGamma_s > \wtGamma_\nu\} = \nu$, which allows us to conclude that $\I_{\{\nu \leq C_{\wtGamma_\nu} < T\}}  R_{C_{\wtGamma_\nu}} = \I_{\{\nu < T\}}  R_{\nu}$.
\end{proof}

We are now ready to prove the main result of this section, which shows that the reduced upper price of a vulnerable European option coincides
with the value process of the constrained optimal stopping problem \eqref{osp1}. Recall that $\barS = S^r(A^o) \cup \{T\}$ where
$S^r(A^o)$ is the right support of $A^o$.

\bt \label{th3.1}
The reduced upper price $V^u :=\esssup_{\,\varphi \in\Phi}V^{\varphi}$ satisfies the equality $V^u=\barV^{\FF}$ where $\barV^{\FF}$ is the value process for the constrained optimal stopping problem \eqref{osp1}, that is, for every $t\in [0,T]$,
\begin{equation}  \label{eq5.38}
V^u_t = \esssup_{\tau\in \bSTFt}\EQ\big[P_{T}\I_{\{\tau=T\}}+R_{\tau}\I_{\{\tau< T\}}\,|\,\cF_t\big]
\end{equation}
where $\bSTFt$ is the class of all $\FF$-stopping times with values in $\barS \cap [t,T]$.
\et

\begin{proof}
As in the proof of Proposition \ref{npro1.15}, for every $n\in\NN$, we consider the unique solution $(\wtY^n,\wtZ^n)\in \cS^2(\FF)\times \cH^2(M)$ to the GBSDE
\begin{align*}  
\wtY^n_t=P_T+\int_{]t,T]} n\big(R_s-\wtY^n_s\big)^+\,d\wtGamma_s-\int_{]t,T]} \wtZ^n_s\,d\whMM_s ,
\end{align*}
and we record that the sequence $(\wtY^n)_{n\in\NN}$ is an increasing sequence of processes converging to the process $\barY$. Furthermore, we know from Proposition \ref{npro1.15} that $\barY=\barV^{\FF}$ where $\barV^{\FF}$ is given by \eqref{osp1}.

To establish the equality $V^u=\barY$, we will first show that $V^u\leq\barY$. For every $n\in \NN$ the subset $\Phi^{n}$
of $\Phi$ is given by $\Phi^{n}:=\{ \varphi\in\Phi : \varphi^{(o)}\leq n-1\}$ so that $\Phi=\cup_{n\in\NN}\,\Phi^{n}$. Since $\wtGamma $ is a
continuous process we deduce from Corollary \ref{cor5.9} that the process $V^{\varphi}$ satisfies the GBSDE under $\QQ$ (see equation \eqref{eq5.36})
\begin{align*}
V^{\varphi}_t=P_T+\int_{]t,T]}(\varphi^{(o)}_s+1)(R_s-V^{\varphi}_s)\,d\wtGamma_s-\int_{]t,T]}Z^{\varphi}_s\,d\whMM_s.
\end{align*}
Let $V^n_t:=\esssup_{\,\varphi \in\Phi^n}V^{\varphi}_t$. From the comparison theorem for the GBSDE (see Proposition 3.1
in \cite{LLR2}) we deduce that $V^{\varphi}\leq \wtY^{n}$ for every $\varphi\in\Phi^{n}$. Consequently, $V^n\leq \wtY^{n}\leq \barY$, which
entails the desired inequality $V^u\leq\barY = \barV^{\FF}$.
%


In the next step, we will establish the inequality $\barY \leq V^u$. We consider the bounded and strictly positive function
\begin{align*}
g(x)  = n\I_{\{R-x>0\}} + \epsilon\left(\I_{\{-1\leq R-x \leq 0\}} + |R-x|^{-1}\I_{\{R-x< -1\}}\right).
\end{align*}
which is a Lipschitz continuous function in $x$ and satisfies $0\leq n(R-x)^+ - (R-x)g(x) \leq \epsilon$. From the comparison theorem for
GBSDEs (Proposition 3.1 in \cite{LLR2}) we obtain, for every $\epsilon \in (0,1)$,
\begin{align*}
\wtY^n_t & = P_T+\int_{]t,T]} n\big(R_s-\wtY^n_s\big)^+\,d\wtGamma_s-\int_{]t,T]} \wtZ^n_s\,d\whMM_s\\
		 & \leq \wtY^{n,\epsilon}_t = P_T+\int_{]t,T]} \big[\big(R_s-\wtY^{n,\epsilon}_s\big)g(\wtY^{n,\epsilon}_s) + \epsilon \big] \,d\wtGamma_s-\int_{]t,T]} \wtZ^{n,\epsilon}_s\,d\whMM_s,
\end{align*}
where the existence of a unique solution $(\wtY^{n,\epsilon}, \wtZ^{n,\epsilon}) \in \mathcal{S}^2\times \mathcal{H}^2(M^\QQ)$ follows from Proposition 3.5 in \cite{LLR2} since $(R-x)g(x) + \epsilon$ is non-negative, Lipschitz and nonincreasing in $x$.

Let us take $\varphi^{n,\epsilon} := g(\wtY^{n,\epsilon})-1 \in \Phi^n$ and examine the reduced value of the vulnerable European option, given by $V^{\varphi^{n,\epsilon}}$, with the hazard process given by $\wtGamma^{\varphi^{n,\epsilon}} =  g(\wtY^{n,\epsilon})\bcd \wtGamma$. For simplicity, in the following, we will write $V^{{n,\epsilon}}$ instead of $V^{\varphi^{n,\epsilon}}$ and by Proposition \ref{pro5.1x} there exists a unique process $Z^{n,\epsilon}\in \mathcal{H}^2(M^\QQ)$ such that
\begin{align*}
V^{n,\epsilon}_t & = P_T+\int_{]t,T]} \big(R_s-V^{{n,\epsilon}}_s \big)g(\wtY^{n,\epsilon}_s)\,d\wtGamma_s-\int_{]t,T]} Z^{n,\epsilon}_s\,d\whMM_s.
\end{align*}
Next, by using the fact that $g$ is positive we note that the generators $f_s^\epsilon(x) :=  \big(R_s-x \big)g(\wtY^{n,\epsilon}_s) + \epsilon$ and $f_s(x) := \big(R_s-x \big)g(\wtY^{n,\epsilon}_s)$ are Lipschitz and nonincreasing in $x$. Then by using the comparison theorem for GBSDE (see Proposition 3.1 in \cite{LLR2}) we obtain
\begin{align*}
0 & \leq \wtY^{n,\epsilon}_t - V^{{n,\epsilon}}_t\\
& = \int_{]t,T]} [f^\epsilon_s(\wtY^{n,\epsilon}_s ) - f_s(V^{{n,\epsilon}}_s)] d\wtGamma_s + \int_{]t,T]}[\wtZ^{n,\epsilon}_s - Z^{n,\epsilon}_s]\, dM_s^\QQ\\
& = \int_{]t,T]} [f^\epsilon_s(\wtY^{n,\epsilon}_s ) - f_s(\wtY^{n,\epsilon}_s)] d\wtGamma_s + \int_{]t,T]} [f_s(\wtY^{n,\epsilon}_s ) - f_s(V^{{n,\epsilon}}_s)] d\wtGamma_s + \int_{]t,T]}[\wtZ^{n,\epsilon}_s - Z^{n,\epsilon}_s]\, dM_s^\QQ\\
& \leq \epsilon(\wtGamma_T- \wtGamma_t) + \int_{]t,T]}[\wtZ^{n,\epsilon}_s - Z^{n,\epsilon}_s]\, dM_s^\QQ
\end{align*}
where in the last inequality we have used the fact that $\wtY^{n,\epsilon} \geq  \wtY^n$ and $f_s$ is nonincreasing in $x$. Finally, by taking the $\cF_t$-conditional expectation with respect to $\QQ$, we obtain
\begin{gather*}
\wtY^{n}_t \leq \wtY^{n,\epsilon}_t \leq V^{n,\epsilon}_t  + \epsilon \EQ[\wtGamma_T|\cF_t]
\end{gather*}
This shows that $\wtY^n\leq V^{n} \leq V^{u}$ and hence $\barY \leq V^u$, which is the desired inequality.
%
%
\end{proof}

\section{Vulnerable Optimal Stopping}      \label{sec4}

For the reader's convenience, we start by recalling the classical version of the Doob-Meyer-Mertens (DMM) decomposition theorem (see Mertens \cite{M1972} and El Karoui \cite{EK1981}). We stress that the filtration $\FF$ is supposed to satisfy the usual conditions of right-continuity and completeness and we refer to Gal'\v{c}uk \cite{G1982} for an extended version of the DMM decomposition where these assumptions about $\FF$ are dropped. For simplicity, we henceforth work under Assumption \ref{ass3.1} (ii), that is, the process $\wtGamma$ is assumed to be continuous and bounded.

\bp \label{pro4.1}
Any strong $\FF$-supermartingale $Y$ of class (D) has the unique  Doob-Meyer-Mertens (DMM) decomposition $Y=N-B^c-B^d-B^g$ where $N$ is a uniformly integrable $\FF$-martingale, $B^c$ is an $\FF$-adapted, nondecreasing, continuous process, $B^d$ is an $\FF$-predictable, nondecreasing, purely discontinuous process, and $B^g$ is an $\FF$-adapted, c\`agl\`ad, nondecreasing, purely discontinuous process.
\ep

\brem \label{rem4.1}
Recall that any l\`adl\`ag, $\FF$-predictable, nondecreasing process $B$ has a unique decomposition $B=B^c+B^d+B^g$ with the properties stated in Proposition \ref{pro4.1}. Observe also that if we denote $A:=B^c+B^d$ and $C:=B^g_+$, then from Proposition \ref{pro4.1} we obtain the DMM decomposition $Y=N-A-C_{-}$, which is encountered in papers on the optimal stopping, e.g., \cite{EK1981,GIOOQ2017,GIOQ2020,KQ2012}. Alternatively, if we set $D=A+C_{-}$, then we obtain the decomposition $Y=N-D$ where $D$ is a nondecreasing, $\FF$-{\it strongly predictable} process, in the sense that $D$ is $\FF$-predictable and $D_{+}$ is $\FF$-optional (see \cite{G1981,G1982}).
\erem
%
%

Let $\cK$ (resp., $\bcK$) denote the class of all c\`adl\`ag, nondecreasing, $\FF$-predictable (resp., l\`adl\`ag, nondecreasing, $\FF$-predictable) processes and let $\STF$ (resp., $\STFp$) stand for the class of all $\FF$-stopping times (resp., $\FF$-predictable stopping times) $\tau$ taking values in $[0,T]$.

In the next result, we adopt the definition of the reflected BSDE from \cite{GIOQ2020} (see Definition 2.3 in \cite{GIOQ2020}).
Recall that we write $\Delta K^d_{\sigma} :=K^d_{\sigma}-K^d_{\sigma-}$ and  $\Delta^+ K^g_{\sigma} := K^g_{\sigma+}-K^g_{\sigma}$.
It is known that the process $\bar{P}$ given by $\bar{P}_t:=\limsup_{s \uparrow t,\,s<t}P_t$ for
all $t \in ]0,T]$ is $\FF$-predictable and left-upper-semicontinuous (see Theorem 90 on page 225 in \cite{DM1975} or \cite{GIOQ2020}).

It is convenient to introduce the $\FF$-optional, bounded reward process $\xi$ given by, for all $t\in [0,T]$,
\begin{equation} \label{eq4.4}
\xi_t=P_t G_t+\int_{]0,t]}R_s\,dA^o_s=P_tG_t+(R\bcd A^o)_t
\end{equation}
and to study the optimal stopping problem $(\xi,\STF)$ under $\QQ$ with the $\FF$-Snell envelope $\breve{\xi}$ given by
\begin{equation*} 
\breve{\xi}_t:=\esssup_{\sigma\in\STFt}\,\EQ[\xi_{\sigma}|\,\cF_t]=\esssup_{\sigma\in\STFt}\,\EQ \big[P_{\sigma}G_{\sigma}+(R\bcd A^o)_{\sigma}\,|\,\cF_t\big].
\end{equation*}

\bp \label{pro4.2}
Let the process $\VV$ be given by
\begin{equation} \label{eq4.3}
\VV_t:=\esssup_{\sigma\in\STFt} G_t^{-1}\,\EQ\Big[P_{\sigma} G_{\sigma}+\int_{\rrb t,\sigma\rrb}R_s\,dA^o_s\,\Big|\,\cF_t\Big].
\end{equation}
If Assumptions \ref{ass2.1} and \ref{ass3.1} (i)--(ii) are satisfied, then there exist an $\FF$-predictable process $Z$ and a nondecreasing, l\`adl\`ag, $\FF$-predictable process $K$ with $K_0=0$ such that the triplet $(\VV,Z,K)\in \cS^2(\FF) \times \cH^2(\whMM)\times \bcK$ satisfies the reflected BSDE
\begin{equation} \label{eq4.6}
\VV_t=P_T+\int_{]t,T]}(R_s-\VV_s)\, d\wtGamma_s +\int_{]t,T]}\wtG^{-1}_sZ_s\,d[\whMM,m]_s-\int_{]t,T]}Z_s\,d\whMM_s+K_T-K_t
\end{equation}
where $\VV \geq P$ and $K$ has the decomposition $K=K^c+K^d+K^g$ where the processes $K^c,K^d$ and $K^g$ satisfy the Skorokhod conditions
\begin{equation}  \label{eq4.7}
\big(\I_{\{\VV_{-}>\bar{P}\}}\bcd K^c\big)_T=0 ,\ \
(\VV_{\sigma-}-\bar{P}_{\sigma}) \Delta K^d_{\sigma}=0, \ \forall\, \sigma \in \STFp,\ \
(\VV_{\sigma}-P_{\sigma})\Delta^+ K^g_{\sigma}=0, \ \forall\, \sigma \in \STF,
\end{equation}
where $\STFp$ is the class of all $\FF$-predictable stopping times with values in $[0,T]$.
\ep

\begin{proof}
We first apply the classical results from El Karoui \cite{EK1981} and Maingueneau \cite{M1978} (see also Kobylanski and Quenez \cite{KQ2012} for an extension) to the optimal stopping problem \eqref{eq4.3}. Recall that $\breve{\xi}$ denotes the Snell envelope of the bounded, $\FF$-optional reward process $\xi$. Hence $\breve{\xi}$ is a bounded, strong $\FF$-supermartingale (hence it is l\`adl\`ag but not necessarily c\`adl\`ag) and the DMM decomposition of $\breve{\xi}$ yields an $\FF$-martingale $N$ and nondecreasing processes $B^c,B^d$ and $B^g$ such that $\breve{\xi}=N-B^c-B^d-B^g$. Notice that $B^r:=B^c+B^d$ is a c\`adl\`ag, nondecreasing and $\FF$-predictable process.  Consequently, the process $\VV$ is l\`adl\`ag and satisfies
\begin{equation} \label{eq4.8}
\VV=G^{-1}\big(\breve{\xi}-(R\bcd A^p)\big) = G^{-1}\big(N-(R\bcd A^p)-B^c-B^d-B^g \big)
\end{equation}
where $B^c,B^d$ and $B^g$ satisfy the Skorokhod conditions
\begin{equation}  \label{eq4.9}
\big( \I_{\{\breve{\xi}_{-}>\bxi\}}\bcd B^c\big)_T=0 ,\ \
(\breve{\xi}_{\sigma-}-\bxi_{\sigma}) \Delta B^d_{\sigma}=0, \ \forall\, \sigma \in \STFp,\ \
(\breve{\xi}_{\sigma}-\xi_{\sigma})\Delta^+ B^g_{\sigma}=0, \ \forall\, \sigma \in \STF .
\end{equation}
By applying Lemma 7.3 in \cite{ALR2022} to \eqref{eq4.8},
we obtain the existence of an $\FF$-predictable process $Z$ such that
\begin{align*}
\VV_t&=P_T+\int_{]t,T]}(R_s-\VV_s)\,\wtGamma_s +\int_{]t,T]}\wtG^{-1}_s Z_s\,d[\whMM,m]_s \\
&\quad -\int_{]t,T]}Z_s\,d\whMM_s+\int_{]t,T]}\wtG^{-1}_s\,dB^r_s+\int_{[t,T[}G^{-1}_s\,dB^g_{s+}.
\end{align*}
If we set
\[
K^c:=\wtG^{-1}\bcd B^c,\quad K^d:=\wtG^{-1}\bcd B^d,\quad K^g:= G^{-1}\bcd B^g_{+} ,
\]
then it suffices to show that the processes $K^c,K^d$ and $K^g$ satisfy the Skorokhod conditions \eqref{eq4.7}. We first observe that $\VV G=\breve{\xi}-(R\bcd A^o)$ and thus $\VV_-G_-=\breve{\xi}_- -(R\bcd A^o)_-$. Furthermore, from \eqref{eq4.4} we obtain $\bxi=\bar{P} G_-+ (R\bcd A^o)_-$. Since the processes $\wtG^{-1}$ and $G^{-1}$ are strictly positive, the conditions stated in \eqref{eq4.9} yield
the Skorokhod conditions \eqref{eq4.7} and thus we conclude that the triplet $(\VV,Z,K)$ is a solution
to the reflected BSDE \eqref{eq4.6}--\eqref{eq4.7}.
\end{proof}

\section{Vulnerable American Options}      \label{sec5}

In this final section, we examine American style options with a possible exogenous termination at a random time $\vt$. Recall that the payoff  process $\whX\in\cOgg$ is given by $\whX:=P\I_{\llb 0,\vt\llb}+R_\vt\I_{\llb\vt,\infty\llb}$ so that for every $\whsigma \in \STG$
\begin{equation} \label{eq6.1}
\whX_{\whsigma}=\whX_{\whsigma\wedge\vt}=P_{\whsigma}\I_\seq{\whsigma<\vt}+R_\vt\I_\seq{\whsigma\geq\vt}.
\end{equation}

\brem
Alternatively, it would be possible to assume that the payoff  process is of the form $\whY:=P\I_{\llb 0,\vt\rrb}+R_\vt\I_{\rrb\vt,\infty\llb}$, which would give, for every $\whsigma \in \STG$,
$$\whY_{\whsigma}=P_{\whsigma}\I_\seq{\whsigma\leq\vt}+R_\vt\I_\seq{\whsigma>\vt}.$$
Notice that this convention  was adopted by Szimayer \cite{S2005} but within the setup studied in \cite{S2005}
the equality $\PP (\whsigma =\vt)=0$ is satisfied for every $\whsigma \in \STG$ and thus the choice of convention is immaterial
since the equality $\PP (\whX_{\whsigma}=\whY_{\whsigma})=1$ holds for every $\whsigma\in\STG$ within the setup of \cite{S2005}.
\erem
%
%

We henceforth work under Assumptions \ref{ass2.1} and \ref{ass3.1} without further mention and we extend Definitions \ref{def3.2} and \ref{def5.1}
to the case of an American option. In particular, we have the following definition of the upper and lower price of a {\it vulnerable American option},
which is represented by a pair $(\whX,\STG)$ where $\whX$ is the payoff received by the holder upon either exercise or default, whichever comes first, and $\STG$ is the class of all exercise times available to the holder of the American option.

\bd \label{def6.1}
The {\it upper price} of the vulnerable American option  $(\whX,\STG)$ is given by, for every $t\in [0,T]$,
\begin{equation} \label{eq6.3}
\whV^u_t:=\esssup_{\QQphi\in\cQ}\esssup_{\whsigma\in\STGt}\EQphi[\whX_{\whsigma}\,|\,\cG_t].
\end{equation}
The {\it lower price} of the vulnerable American option  $(\whX,\STG)$ is given by, for every $t\in [0,T]$,
\begin{equation} \label{eq6.4}
\whV^l_t:=\essinf_{\QQphi\in\cQ}\esssup_{\whsigma\in\STGt}\EQphi[\whX_{\whsigma}\,|\,\cG_t].
\end{equation}
\ed

If we denote, for any fixed $\QQphi\in\cQ$,
\begin{align*}
\whV_t^{\QQphi}=\esssup_{\whsigma\in\STGt}\EQphi[\whX_{\whsigma}\,|\,\cG_t],
\end{align*}
then the upper and lower price can be represented as follows
\begin{align*}
\whV^u_t=\esssup_{\QQphi\in\cQ}\whV_t^{\QQphi}=\esssup_{\QQphi\in\cQ} \big[\I_\seq{\vt \leq t}R_{\vt}+\I_\seq{\vt>t}V_t^{\QQphi}\big]=
 \I_\seq{\vt \leq t}R_{\vt}+\I_\seq{\vt>t} \esssup_{\varphi\in\Phi} V_t^{\varphi}
\end{align*}
and
\begin{align*}
\whV^l_t=\essinf_{\QQphi\in\cQ}\whV_t^{\QQphi}=\essinf_{\QQphi\in\cQ}\big[\I_\seq{\vt \leq t}R_{\vt}+\I_\seq{\vt>t}V_t^{\QQphi}\big]=
 \I_\seq{\vt \leq t}R_{\vt}+\I_\seq{\vt>t} \essinf_{\varphi\in\Phi} V_t^{\varphi}
\end{align*}
where $V_t^{\varphi}$ is called the $\FF$-{\it reduced price under} $\QQphi\in\cQ$ of the vulnerable American option. It is also clear that, for any $\varphi\in \Phi$,
\begin{equation} \label{eq6.41}
\VV_t^{\varphi}:=\esssup_{\sigma\in\STFt} (H^{\varphi}_t)^{-1}\,\EP\Big[P_{\sigma} H^{\varphi}_{\sigma}+\int_{\rrb t,\sigma\rrb}R_s\,dD^{\varphi}_s\,\Big|\,\cF_t\Big].
\end{equation}

\bd \label{def6.2}
The {\it $\FF$-reduced upper price} (resp., {\it $\FF$-reduced lower price}) of the vulnerable American option  $(\whX,\STG)$
is denoted by $V^u$ (resp., $V^l$) and is given by, for every $t\in [0,T]$,
\begin{align*}
V_t^u:=\esssup_{\varphi\in\Phi}V_t^{\varphi}, \qquad V_t^l:=\essinf_{\varphi\in\Phi}V_t^{\varphi}.
\end{align*}
\ed

It is clear that $V^u$ and $V^l$ are $\FF$-adapted processes and the equalities $V_t^u=\whV^u_t$ and $V_t^l=\whV^l_t$ are satisfied
on the event $\seq{\vt>t}$ for every $t\in[0,T]$. Furthermore, we define
\begin{align*}
V_t^{u,n}:=\esssup_{\varphi\in\Phi^n}V_t^{\varphi},\qquad V_t^{l,n}:=\essinf_{\varphi\in\Phi^n}V_t^{\varphi}
\end{align*}
and we observe that $V_t^{u,n+1}\geq V_t^{u,n}$ and $V^u_t=\lim_{n\to\infty}V^{u,n}_t$  (resp., $V_t^{l,n+1}\leq V_t^{l,n}$ and
$V^l_t=\lim_{n\to\infty}V^{l,n}_t$). The following crucial lemma shows that the processes $V^{u,n}$ and $V^{l,n}$ are solutions
to reflected GBSDEs \eqref{eq6.51} and \eqref{eq6.52}, respectively.

\bl \label{lem4.2}
(i) Let $(Y^n,Z^n,K^n)$ be a unique solution in $\cS^2(\FF) \times \cH^2(M)\times \bcK$ to the reflected GBSDE
\begin{align} \label{eq6.51}
Y^n_{\tau}=P_T-\int_{\rrb\tau,T\rrb} Z^n_s\,d\whMM_s+\int_{\rrb\tau,T\rrb} n(R_s-Y^n_s)^+\,d\wtGamma_s+K^n_T-K^n_{\tau}
\end{align}
where $Y^n \geq P$ and a nondecreasing, l\`adl\`ag, $\FF$-predictable process $K^n$ satisfies the Skorokhod conditions
\eqref{eq4.7} with $Y^n$ and the lower obstacle $P$. Then $Y^{n+1} \geq Y^{n}$ and $\VV^{u,n}=Y^n$ for every $n\in\NN$. \\
(ii)  Let $(\wtY^n,\wtZ^n,\wt{K}^n)$ be a unique solution in $\cS^2(\FF) \times \cH^2(M)\times \bcK$ to the reflected GBSDE
\begin{align}  \label{eq6.52}
\wtY^n_{\tau}=P_T-\int_{\rrb\tau,T\rrb} \wtZ^n_s\,d\whMM_s-\int_{\rrb\tau,T\rrb} n(\wtY^n_s-R_s)^+\,d\wtGamma_s+\wt{K}^n_T-\wt{K}^n_{\tau}
\end{align}
where $\wtY^n \geq P$ and a nondecreasing, l\`adl\`ag, $\FF$-predictable process $\wt{K}^n$ satisfies the Skorokhod conditions \eqref{eq4.7}
with $\wtY^n$ and the lower obstacle $P$. Then $\wtY^{n+1} \leq \wtY^{n}$ and $\VV^{l,n}=\wtY^n$ for every $n\in\NN$.
\el


\begin{proof} (i)
By performing computations similar to \eqref{eq3.18}, we obtain from \eqref{eq6.41}
\begin{align*}
\VV_t^{\varphi} & =\esssup_{\sigma\in\STFt} (H^{\varphi}_t)^{-1}\,\EP\Big[P_{\sigma} H^{\varphi}_{\sigma}+\int_{\rrb t,\sigma\rrb}R_s\,dD^{\varphi}_s\,\Big|\,\cF_t\Big] \\
				& = \esssup_{\sigma\in\STFt} (\cE_t(-\wtGamma^\varphi))^{-1}\,\EQ\Big[P_{\sigma} \cE_{\sigma}(-\wtGamma^\varphi)+\int_{\rrb t,\sigma\rrb}R_s\,d\cE_{s}(-\wtGamma^\varphi)\,\Big|\,\cF_t\Big].
\end{align*}
By applying Proposition \ref{pro4.2} from the appendix (or Proposition 5.1 in Aksamit et al. \cite{ALR2022} with $\mathbb{P} = \QQ$, $G = \cE(-\wtGamma^\varphi),\, m = 1$ and $A^o = 1-\cE(-\wtGamma^\varphi)$) we obtain the GBSDE
\begin{equation} \label{eq6.53}
\VV_{\tau}^{\varphi}=P_T-\int_{\rrb\tau,T\rrb}Z^{\varphi}_s\,d\whMM_s+\int_{\rrb\tau,T\rrb} (\varphi^{(o)}_s+1)(R_s-\VV_s^{\varphi})\,d\wtGamma_s+K_T-K_{\tau}.
\end{equation}
Recall that for every $n\in\NN$ the subset $\Phi^n$ of $\Phi$ is given by $\Phi^n:=\{ \varphi\in\Phi : \varphi^{(o)}\leq n-1\}$ and $\VV_t^{u,n}:=\esssup_{\varphi\in\Phi^n}\VV_t^{\varphi}$. On the one hand, from the comparison theorem for the reflected GBSDE (see Proposition 4.2 in \cite{LLR2}) we have that $\VV^{\varphi}\leq Y^n$ for every $\varphi\in\Phi^n$ and thus $\VV^{u,n}\leq Y^n$.

On the other hand, we will show that $\VV^{u,n}\geq Y^n$ for every $n \in \NN$. To this end, we fix $n\in \NN$ and $0<\varepsilon<1$ and we consider the reflected GBSDE
\begin{align} \label{eq6.51x}
Y^{n,\varepsilon}_{\tau}=P_T-\int_{\rrb\tau,T\rrb} Z^{n,\varepsilon}_s\,d\whMM_s+\int_{\rrb\tau,T\rrb} g^{n,\varepsilon} (Y^{n,\varepsilon}_s)\,d\wtGamma_s+K^{n,\varepsilon}_T-K^{n,\varepsilon}_{\tau}
\end{align}
with the lower obstacle $P$ where the generator $g^{n,\varepsilon}$ is given by
\begin{align*}
g^{n,\varepsilon}(Y^{n,\varepsilon})=n(R-Y^{n,\varepsilon})^+ + \varepsilon (R-Y^{n,\varepsilon})\I_{\{-1\leq R-Y^{n,\varepsilon}\leq 0\}}
- \varepsilon \I_{\{R-Y^{n,\varepsilon}<-1\}}\leq n(R-Y^{n,\varepsilon})^+
\end{align*}
and $Y^{n,\varepsilon} \geq P$. We observe that the reflected GBSDE \eqref{eq6.51x} has a unique solution, which satisfies
$Y^n - C \varepsilon  \leq Y^{n,\varepsilon}\leq Y^n$ for every $0<\varepsilon<1$ and a constant $C$ independent of $\varepsilon$.
This follows from Section 4.2 in \cite{LLR2} where it is shown that if $|n(R-Y^{n,\varepsilon})^+ - g^{n,\varepsilon}(Y^{n,\varepsilon})|\leq \varepsilon$
and the process $\wtGamma$ is bounded, then there exists a constant $C$ independent of $\varepsilon$
(notice that $C$ may depend on a Lipschitz constant $n$ in the generator of a reflected GBSDE but $n$ is here fixed)
such that
\begin{align*}
\sup_{t\in [0,T]} |Y^{n,\varepsilon}_t -Y^{n}_t| \leq C \varepsilon .
\end{align*}
Then, for a unique solution $Y^{n,\varepsilon}$ to the GBSDE \eqref{eq6.51x}, we define $\varphi^{n,\varepsilon} \in\Phi^{n}$
\begin{align*}
\varphi^{n,\varepsilon}:=n\I_{\{R-Y^{n,\varepsilon}>0\}}+\varepsilon \I_{\{-1\leq R-Y^{n,\varepsilon}\leq 0\}}-
\varepsilon (R-Y^{n,\varepsilon})^{-1}\I_{\{R-Y^{n,\varepsilon}\leq -1\}}-1
\end{align*}
and we consider the reflected GBSDE
\begin{align} \label{eq6.51y}
\VV_{\tau}^{n,\varepsilon}=P_T-\int_{\rrb\tau,T\rrb}Z^{n,\varepsilon}_s\,d\whMM_s+\int_{\rrb\tau,T\rrb} (\varphi^{n,\varepsilon}_s+1)(R_s-\VV_s^{n,\varepsilon})\,d\wtGamma_s
+K^{n,\varepsilon}_T-K^{n,\varepsilon}_{\tau}
\end{align}
with the lower obstacle $P$. It is clear that equation \eqref{eq6.51y} has a unique solution $\VV^{n,\varepsilon}$. Furthermore, a comparison of
equations \eqref{eq6.51x} and \eqref{eq6.51y} makes it clear that the equality $\VV^{n,\varepsilon}=Y^{n,\varepsilon}$ holds for every $n\in\NN$ and $0\leq \varepsilon<1$. 
Since $\VV^{n,\varepsilon}=Y^{n,\varepsilon}$ and $Y^{n,\varepsilon}\geq Y^n - C \varepsilon$ where $\varepsilon$ is arbitrarily small
we conclude that $\VV_t^{u,n}:=\esssup_{\varphi\in\Phi^{n}}\VV_t^{\varphi} \geq Y^n_t $.

\noindent (ii) The arguments are similar and also hinge on the comparison theorem for reflected GBSDE given in Proposition 4.2 in \cite{LLR2}. We now define
$\VV_t^{l,n}:=\essinf_{\varphi\in\Phi^n}\VV_t^{\varphi}$ observe that \eqref{eq6.53} is clearly equivalent to
\begin{equation*}
\VV_{\tau}^{\varphi}=P_T-\int_{\rrb\tau,T\rrb}Z^{\varphi}_s\,d\whMM_s-\int_{\rrb\tau,T\rrb}(\varphi^{(o)}_s+1)(\VV_s^{\varphi}-R_s)\,d\wtGamma_s+K_T-K_{\tau}.
\end{equation*}
Therefore, it suffices to consider the process $\wtY^n$ where the triplet $(\wtY^n,\wtZ^n,\wt{K}^n)$ is a unique solution to \eqref{eq6.52}
and use the comparison theorem for reflected BSDEs from \cite{LLR2} to deduce that the inequality $\VV^{\varphi}\geq \wtY^n$ holds for every $\varphi\in\Phi^n$ and thus $\VV^{l,n}\geq \wtY^n$. Then the inequality $\VV^{l,n}\leq \wtY^n$ can be obtained by arguing along similar lines as in part (i).
\end{proof}

\subsection{Issuer's Duality Theorem}          \label{sec6.1}

Let us recall that $\barS = S^r\cup\{T\}$ where $S^r=S^r(A^o)$ is the right support of the process $A^o$, that is, $S^r(A^o):=\{ t\in \RR_+ : \forall\,\varepsilon>0 \ A^o_{t+\varepsilon}-A^o_t>0\}$. The first main result for vulnerable American options shows that the upper price is a solution to
the optimal stopping problem \eqref{eq6.5}.

\bt \label{th6.1}
Let the process $P$ (resp., $R$) be l\`adl\`ag and right-upper-semicontinuous (resp., right-continuous).
Then the $\FF$-reduced upper price $V^u$ satisfies, for every $t\in [0,T]$,
\begin{equation} \label{eq6.5}
V^u_t=\esssup_{\sigma\in\STFt}\EQ\big[\zeta^u_\sigma\,|\,\cF_t\big] 
\end{equation}
where the process $\zeta^u$ is given by $\zeta^u_t := (P\vee R\I_{\barS})_t\I_{\{t<T\}}+P_T \I_{\{t=T\}}$.
\et

\begin{proof}
We use a similar method as in the proof of Theorem \ref{th3.1}. In view of part (i) in Lemma \ref{lem4.2}, we have that $\VV^{u,n}=Y^n$
where $(Y^n,Z^n,K^n)$ is a unique solution in $\cS^2(\FF) \times \cH^2(M)\times \bcK$ to the reflected GBSDE (see Proposition 4.4 in \cite{LLR2})
\begin{align} \label{eq6.6}
Y^n_{\tau}=P_T-\int_{\rrb\tau,T\rrb} Z^n_s\,d\whMM_s+\int_{\rrb\tau,T\rrb} n(R_s-Y^n_s)^+\,d\wtGamma_s+K^n_T-K^n_{\tau}
\end{align}
with the lower obstacle $P$ and the Skorokhod conditions satisfied by $Y^n$ and an $\FF$-adapted, l\`adl\`ag, nondecreasing process $K^n$. We note that, in view of Corollary 4.1 in \cite{LLR2}, the sequence $Y^n$ of processes is monotonically increasing as $n\rightarrow \infty$ and the equalities $V^u=\lim_{n\rightarrow \infty}Y^n=\lim_{n\rightarrow \infty}V^{u,n}$ hold.

\noindent {\it Step 1.}  Our first goal is to show that, for every $n\in\NN$,
\begin{align}  \label{eq6.6a}
V^{u,n}_t=Y^n_t=\esssup_{\sigma\in\STFt}\EQ\big[Y^n_\sigma \wedge \zeta^u_\sigma\,|\,\cF_t\big]
         =\esssup_{\sigma\in\STFt}\EQ\big[V^{u,n}_\sigma \wedge \zeta^u_\sigma\,|\,\cF_t\big].
\end{align}
To prove \eqref{eq6.6a}, we fix $n$ and we note that $Y^n$ is an $\FF$-supermartingale and thus, for every $\sigma\in\STFt$,
\begin{align} \label{eq6.6b}
Y^n_t \geq\EQ\big[Y^n_\sigma\,|\,\cF_t\big]\geq \EQ\big[Y^n_\sigma \wedge \zeta^u_\sigma\,|\,\cF_t\big]
\end{align}
where $\zeta^u_t := (P\vee R\I_{\barS})_t\I_{\{t<T\}}+P_T \I_{\{t=T\}}$.

To show the reverse inequality, we fix $t\in [0,T[$ and we
define $\nu =\sigma_t^n\wedge\tau_t^n \in \STFt$ where for an arbitrary $\delta>0$ we define  (by convention, $\inf \emptyset =T$)
\begin{align*}
\sigma_t^n:=\inf\{s\in [t,T]: Y^n_s \leq P_s + \varepsilon \}, \quad \tau_t^n:=\inf\{s\in [t,T]:\wtK^n_s-\wtK^n_t>0\},\
\end{align*}
where $\varepsilon :=0.5(Y^n_t-P_t)\delta $ and the continuous, nondecreasing process $\wtK^n$ is given by $\wtK^n_t:=\int_0^t n(R_s-Y^n_s)^+\,d\wtGamma_s$. We will check that $Y^n_\nu = Y^n_\nu\wedge ( (P+\varepsilon)\vee R\I_{\barS})_\nu$ on the event $\{\nu <T\} =\{\sigma_t^n\leq \tau_t^n<T\} \cup \{\tau_t^n< \sigma_t^n\}=E_1 \cup E_2$. It is obvious that $Y^n_\nu = P_T$ on the event $E_3:=\{\nu =T\}$.

On the event $E_1=\{t<\sigma_t^n\leq \tau_t^n<T\}$, we have $Y^n_{\sigma_t^n-}-P_{\sigma_t^n-} \geq \varepsilon$ and thus $\Delta K^{n,d}_{\sigma_t^n}=0$, which implies that $Y^n$ is an $\FF$-martingale on $\llb t,\sigma_t^n\rrb$. Furthermore, if $\Delta^+ K^{n,g}_{\sigma_t^n}>0$, then the Skorokhod condition
gives $Y^n_{\sigma_t^n}=P_{\sigma_t^n}$ and if $\Delta^+ K^{n,g}_{\sigma_t^n}=0$, then $Y^n$ is continuous at $\sigma_t^n$ and  $P_{\sigma_t^n}\leq
Y^n_{\sigma_t^n}\leq P_{\sigma_t^n}+\varepsilon$ since the process $P$ is assumed to be right-upper-semicontinuous. We conclude that on $E_1$
we have $Y^n_\nu = Y^n_\nu  \wedge (P_\nu + \varepsilon) = Y^n_\nu\wedge ( (P+\varepsilon) \vee R\I_{\barS})_\nu$ where the second equality is a trivial consequence of the first one.

On the event $E_2=\{\tau_t^n < \sigma_t^n\}$, the process $Y^n$ is right-continuous at $\tau_t^n$ and hence from the definition of $\tau_t^n$ we obtain $Y^n_{\tau_t^n}= Y^n_{\tau_t^n+} \leq R_{\tau_t^n+} = R_{\tau_t^n}$ where the inequality follows from the right-continuity of $R$.
We note also that the $\FF$-stopping time $\tau_t^n$ has values in $\barS$ so that $R_{\tau_t^n}=(R\I_{\barS})_{\tau_t^n}$ and thus we have $Y^n_\nu = Y^n_\nu  \wedge (R\I_{\barS})_\nu = Y^n_\nu\wedge ((P+\varepsilon)\vee R\I_{\barS})_\nu$ on $E_2$ where the second equality is obvious.  It is also clear that $Y^n$ is an $\FF$-martingale on $\llb t,\tau_t^n\rrb$ since the continuous, nondecreasing process $\wt K^n$ and the nondecreasing process $K^n$ are constant on that interval.

Recall that $\varepsilon=0.5(Y^n_t-P_t)\delta $ and the processes $Y^n$ and $P$ are bounded so that $\varepsilon \leq c \delta$ for some constant $c$. Let us denote $\zeta^{\varepsilon}_t:=((P+\varepsilon)\vee R\I_{\barS})_t\I_{\{t<T\}}+P_T\I_{\{t=T\}}$. Since $Y^n$ is an $\FF$-martingale on $\llb t,\nu \rrb$ we have
\begin{align} \label{eq6.6c}
Y^n_t=\EQ\big[Y^n_\nu\,|\,\cF_t\big]= \EQ\big[Y^n_\nu \wedge \zeta^{\varepsilon}_\nu\,|\,\cF_t\big]
 \leq \EQ\big[Y^n_\nu \wedge \zeta^u_\nu\,|\,\cF_t\big] + c\delta \leq \EQ\big[Y^n_\nu \wedge \zeta^u_\nu\,|\,\cF_t\big]
\end{align}
where the last inequality holds since $\delta$ is any positive number. By combining \eqref{eq6.6b} with \eqref{eq6.6c} we conclude that
\eqref{eq6.6a} is satisfied for every $n\in\NN$.

\noindent {\it Step 2.} We are now ready to show that  \eqref{eq6.5} is valid. For any $\tau \in \bSTFt$ equation \eqref{eq6.6} gives
\begin{align*}
Y^n_{\tau} & = P_{T}-\int_{\rrb\tau,T\rrb} Z^n_s\,d\whMM_s+\int_{\rrb \tau,T\rrb} n(R_s-Y^n_s)^+\,d\wtGamma_s+K_{T}-K_{\tau}.
\end{align*}
and, by the comparison theorem for GBSDEs (see Proposition 4.2 in \cite{LLR2}) on the interval $\llb \tau, T\rrb$,
we see that $Y^n \geq \hat Y^n \geq \barY^n $ where $(\hat Y^n,\hat Z^n,\hat K^n)$ and $(\barY^n,\barZ^n)$ solve the following linear
RBSDE and BSDE, respectively,
\begin{align*}
\hat Y_\tau^n&=P_T-\int_{\rrb\tau,T\rrb}\hat Z^n_s\,d\whMM_s+\int_{\rrb \tau,T\rrb}n(R_s-\hat Y^n_s)\,d\wtGamma_s+\hat K^n_T-\hat K^n_\tau\\
&=P_T-\int_{\rrb\tau,T\rrb} \hat Z^n_s\,d\whMM_s+ \int_{\rrb \tau, T\rrb} n(R_s- \hat Y^n_s)\,d\wtGamma_s+\hat K^n_T-\hat K^n_\tau\\
&\geq  P_T-\int_{\rrb\tau,T\rrb} \barZ^n_s\,d\whMM_s+ \int_{\rrb \tau, T\rrb} n(R_s- \barY^n_s)\,d\wtGamma_s = \barY^n_{\tau}
\end{align*}
where the inequality holds since the generator $n(R_s-y)$ is linear and the process $\hat K$ is nondecreasing on $\llb \tau, T\rrb$.
Then by solving the linear GBSDE or using Corollary \ref{cor5.9}, we obtain
\begin{align*}
\barY_\tau^n&=\EQ\big[P_{T}\cE_{\tau,T}(-\wtGamma^n)+(\I_{\rrb \tau,T\rrb}R\cE_{\tau,\cdot}(-\wtGamma^n)\bcd \wtGamma^n)_{T}\, | \cF_\tau\big]	.
\end{align*}
The quantity $\cE_{\tau,T}(-\wtGamma^n)$ converges to $\I_{\{\tau=T\}}$ as $n\rightarrow \infty$ and thus, by virtue of Lemma \ref{lem3.4x} and the right continuity of $R$, we obtain
\begin{align}
V^u_\tau=\lim_{n\rightarrow \infty}Y^n_{\tau} \geq \lim_{n \rightarrow \infty} \barY^n_{\tau}=P_{\tau}\I_{\{\tau= \sigma\}}+R_{\tau}\I_{\{\tau< \sigma\}}. \label{VR1}
\end{align}
Using the fact that $V^u\geq \barY^n \geq 0$ and $V^u \geq P$, we deduce from \eqref{VR1} that for any stopping time $\sigma\in\STFt$ we have, on the event $\{\sigma < T\}$,
\begin{gather*}
V^u_\sigma \geq P_{\sigma} \vee (R\I_{\barS})_\sigma
\end{gather*}
and, obviously, $V^u_T = P_T$ on the event $\{\sigma = T\}$. We deduce that
\begin{align*}
V^u_\sigma \geq P_{T}\I_{\{\sigma = T\}}+P_{\sigma} \vee (R\I_{\barS})_\sigma  \I_{\{\sigma < T\}} = \zeta^u_{\sigma}.
\end{align*}
Since we clearly have $Y^n_t \leq \esssup_{\sigma\in\STFt}\EQ\big[\zeta^u_\sigma\,|\,\cF_t\big]$, it suffices to show that
\begin{gather*}
 V^u_t  \geq  \esssup_{\sigma \in\STFt}\EQ\big[\zeta^u_\sigma\,|\,\cF_t\big].
\end{gather*}
The above inequality follows from the observation that, by monotone convergence theorem, $V^u$ is a supermartingale dominating the payoff $\zeta^u$ and the minimality property of the Snell envelope. Hence we have shown that \eqref{eq6.5} is satisfied.
\end{proof}

%

\subsection{Holder's Duality Theorem}             \label{sec6.2}

Our next goal is to prove the duality result for the holder's problem. As in Szimayer \cite{S2005},
we will show that the holder's price can be conveniently represented in terms of a particular zero-sum two-person game.
In the statement of Theorem \ref{th6.2} the stopping time $\sigma\in \STFt$ can be interpreted as the $\FF$-reduction
of the exercise time of the holder of a vulnerable American option, whereas the $\FF$-stopping time $\tau\in\bSTFt$ with values in $\barS\cap[t,T]$ is formally associated with the random time $\vartheta$.

\bt  \label{th6.2}
Let the process $P$ (resp., $R$) be right-upper-semicontinuous (resp., right-continuous). If the inequality $P\leq R$ holds on
the set $\barS$, then the $\FF$-reduced lower price equals, for every $t\in [0,T]$,
\begin{equation} \label{eq6.7}
V^l_t=\essinf_{\tau\in\bSTFt}\esssup_{\sigma\in\STFt}\EQ[X^l(\sigma,\tau)\,|\,\cF_t]=
\esssup_{\sigma\in\STFt}\essinf_{\tau\in\bSTFt}\EQ[X^l(\sigma,\tau)\,|\,\cF_t]
\end{equation}
where
\begin{equation} \label{eq6.8}
X^l(\sigma ,\tau ):=P_{\sigma}\I_\seq{\tau > \sigma}+(P\vee R)_\tau \I_\seq{\tau \leq \sigma}.
\end{equation}
\et

\begin{proof}
In view of part (ii) in Lemma \ref{lem4.2}, the process $\VV^{l,n}$ is equal to the process $\wtY^n$ where the triplet
$(\wtY^n,\wtZ^n,\wt{K}^n)$ is a unique solution  in $\cS^2(\FF) \times \cH^2(M)\times \bcK$ to the reflected GBSDE (see Proposition 4.4 in \cite{LLR2})
\begin{align}  \label{eq6.9}
\wtY^n_{\tau}=P_T-\int_{\rrb\tau,T\rrb} \wtZ^n_s\,d\whMM_s-\int_{\rrb\tau,T\rrb} n(\wtY^n_s-R_s)^+\,d\wtGamma_s+\wt{K}^n_T-\wt{K}^n_{\tau}
\end{align}
where $\wtY^n \geq P$ and the Skorokhod conditions are satisfied by an $\FF$-adapted, nondecreasing process $\wt{K}^n$.
Observe that, as opposed to the maximizer's case, the recovery process $R$ is acting here similarly to an upper obstacle for $\wtY^n$.
We note that the sequence $\wtY^n$ of processes is monotonically decreasing as $n\rightarrow \infty$ (see Corollary 4.1 in \cite{LLR2}) and the equalities $V^l=\lim_{n\rightarrow \infty}\wtY^n=\lim_{n\rightarrow \infty}V^{l,n}$ hold.

\noindent {\it Step 1.} We will first prove that
\begin{align} \label{eq6.11}
V^l_t \geq \essinf_{\tau\in\bSTFt}\esssup_{\sigma\in\STFt}\EQ\big[P_{\sigma}\I_\seq{\tau>\sigma}
+(P\vee R)_\tau \I_\seq{\tau \leq \sigma}\,|\,\cF_t\big].
\end{align}
To establish \eqref{eq6.11}, for any fixed $t$ and $n$, we define $\bar{\tau}^n_t:= \inf\{s\in [t,T]:\wt{L}^n_s-\wt{L}^n_t > 0\}$ where $\wt{L}^n_t:=\int_0^t n(\wtY^n_s-R_s)^+\,d\wtGamma_s$. Since the process $\wt{L}^n$ is continuous, the graph of the stopping time $\bar{\tau}^n_t$ is contained in $\barS \cap [t,T]$ and thus $\bar{\tau}^n_t \in\bSTFt$. Suppose, on the contrary, that $\bar{\tau}^n_t \not \in\bSTFt$. Then the event $\{\bar\tau^n_t < T\}\cap \{\bar\tau^n_t \in \barS^c\}$ has a positive probability and for any fixed $\omega\in \{\bar\tau^n_t < T\}\cap \{\bar\tau^n_t \in \barS^c\} $ there exists $\delta = \delta (\omega) > 0$ such that $\wtGamma_{\bar\tau^n_t + \delta} =\wtGamma_{\bar\tau^n_t}$. However, this contradicts the definition of $\bar\tau^n_t$ since $\wt{L}^n$ is absolutely continuous with respect to $\wt\Gamma$ and thus $\wt{L}^n_{\bar\tau^n_t + \delta} =\wt{L}^n_{\bar\tau^n_t}$.

From the continuity of $\wtGamma$ we obtain $\wtY^n_{\bar{\tau}^n_t+} \geq \liminf_{s\downarrow \bar\tau^n_t}R_{s}$ on $\{\bar{\tau}^n_t<T\}$ and, consequently, using also  \eqref{eq6.9} and the right-continuity of $R$ we deduce that $\wtY^n_{\bar{\tau}^n_t} = \wtY^n_{\bar{\tau}^n_t+}+
\Delta^+ \wt{K}^{n,g}_{\bar{\tau}^n_t}\geq \wtY^n_{\bar{\tau}^n_t+} \geq R_{\bar{\tau}^n_t}$.
In addition, we have $\wtY^n_{\bar{\tau}^n_t} \geq P_{\bar{\tau}^n_t}$ since $(\wtY^n,\wtZ^n,\wtK^n)$ solves the reflected GBSDE \eqref{eq6.9}.
We conclude that $\wtY^n_{\bar{\tau}^n_t} \geq (P\vee R)_{\bar{\tau}^n_t}$ on $\{\bar{\tau}^n_t<T\}$ and, manifestly, $\wtY^n_T=P_T$.

We now take an arbitrary stopping time $\sigma \in \STFt$ and define $\nu :=\bar{\tau}^n_t\wedge\sigma$ so that $\wtY^n$
is a strong $\FF$-supermartingale on $\llb t,\nu\rrb$ since $\wt{L}^n_{\nu}=\wt{L}^n_t$. Then $\wtY^n_\nu \geq P_\nu$ on $E_1:=\{\bar{\tau}^n_t\geq \sigma\}$ and
$\wtY^n_{\nu}\geq (P\vee R)_{\nu}$ on $E_2:=\{\bar{\tau}^n_t < \sigma\}$. Consequently, for any $\sigma\in\STFt$,
\begin{align}\label{eq6.11g}
\wtY^n_t\geq \EQ[ \wtY^n_{\nu}\,|\,\cF_t]\geq \EQ\big[ P_\sigma\I_\seq{\bar{\tau}^n_t>  \sigma}
+(P\vee R)_{\bar{\tau}^n_t}\I_\seq{\bar{\tau}^n_t \leq \sigma}\,|\,\cF_t\big]
\end{align}
from which we deduce that
\begin{align*}
\wtY^n_t & \geq \esssup_{\sigma\in\STFt} \EQ\big[ P_\sigma\I_\seq{\bar{\tau}^n_t>  \sigma}
+(P\vee R)_{\bar{\tau}^n_t}\I_\seq{\bar{\tau}^n_t \leq \sigma}\,|\,\cF_t\big]\\
& \geq \essinf_{\tau\in\bSTFt}\esssup_{\sigma\in\STFt} \EQ\big[ P_\sigma\I_\seq{\tau>  \sigma}
+(P\vee R)_{\tau}\I_\seq{\tau \leq \sigma}\,|\,\cF_t\big].
\end{align*}

Finally, the sequence $\wtY^n=V^{l,n}$ is decreasing and $V^l=\lim_{n\rightarrow \infty}\wtY^n=\lim_{n\rightarrow \infty}V^{l,n}$ so that we obtain \eqref{eq6.11}.

\noindent {\it Step 2.}  In this step, we will establish the inequality
\begin{align} \label{eq6.12}
V^l_t \leq \esssup_{\sigma\in\STFt}\essinf_{\tau\in\bSTFt}\EQ\big[P_{\sigma}\I_\seq{\tau > \sigma }
+(P\vee R)_\tau\I_\seq{\tau \leq \sigma}\,|\,\cF_t\big]
\end{align}
by showing that, for any $\varepsilon > 0$, there exists $\bar{\sigma}_t\in\STFt$, which may depend on $\varepsilon$, such that for an arbitrary $\tau\in\bSTFt$ we have
\begin{align} \label{eq6.11b}
V^l_t \leq \EQ\big[P_{\bar{\sigma}_t}\I_\seq{\tau > \bar{\sigma}_t}+(P\vee R)_\tau \I_\seq{\tau \leq \bar{\sigma}_t}\,|\,\cF_t\big] + \varepsilon.
\end{align}

For a fixed $t$ and $\varepsilon>0$, we define $\bar\sigma^n_t := \inf \{ s\in [t,T]: \wtY^n_s \leq P_s + \varepsilon\}$. Recall that the sequence $\wtY^n$ is monotonically decreasing as $n\rightarrow \infty$ and $V^l=\lim_{n\rightarrow \infty}\wtY^n=\lim_{n\rightarrow \infty}V^{l,n}$ so that $\bar\sigma^n_t \geq \bar\sigma^{n+1}_t$. We define an $\FF$-stopping time $\bar\sigma_t:= \lim_{n\rightarrow \infty} \bar\sigma^n_t$.  From the lower bound in \eqref{eq6.11g} we know that $\wtY^n_t \geq 0$, while the comparison theorem for reflected GBSDEs gives, for every $n\in\NN$,
\begin{align} \label{eq6.11x}
V^l_t \leq \wtY^n_t\leq X_t=P_T-\int_{]t,T]} Z_s\,d\whMM_s+K_T-K_t=\esssup_{\tau \in \STFt} \EQ[P_\tau\,|\,\cF_t] \leq c_P
\end{align}
where $(X,Z,K)$ is a solution to the reflected BSDE implicit in \eqref{eq6.11x} with the lower obstacle $P$ (see Section 4 in \cite{LLR2}) and thus the second equality is due to the well-known relationship between a solution to the reflected BSDE and the value of an optimal stopping problem.

From the assumption that $P$ is right-upper-semicontinuous we deduce that $\wtY^n_{\bar\sigma^n_t} \leq P_{\bar\sigma^n_t} + \varepsilon$ where the inequality is trivially satisfied on the event $\{\bar\sigma^n_t = T\}$. Since $\wtY^n$ is a l\`adl\`ag process, in principle,  it would be possible to have sample paths such that $\wtY^n_{\bar{\sigma}^n_t} > P_{\bar\sigma^n_t}+\varepsilon$ and there exists $\delta >0$ such that $\wtY^n \leq P+\varepsilon$ on $]\bar\sigma^n_t , \bar\sigma^n_t +\delta]$. However, by the right-upper-semicontinuity of $P$, this would imply that $\wtY^n_{\bar\sigma^n_t+} \leq P_{\bar\sigma^n_t}+\varepsilon$ and $\Delta^+ \wtK^{n,g}_{\bar\sigma^n_t}>0$. This in turn would lead to a contradiction since, from the Skorokhod condition for $\wtK^{n,g}$, the inequality $\Delta^+ \wtK^{n,g}_{\bar\sigma^n_t}>0$ implies that $\wtY^n_{\bar\sigma^n_t} = P_{\bar\sigma^n_t} < P_{\bar\sigma^n_t} + \varepsilon < \wtY^n_{\bar\sigma^n_t}$. In view of this, we conclude that $\wtY^n > \wtY^n -\varepsilon> P$ on $\llb t,\bar\sigma^n_t \llb$ and $\wtY^n_- > \wtY^n_- -\varepsilon\geq  \bar P$ on $\rrb t,\bar\sigma^n_t\rrb$ which, together with the Skorokhod condition satisfied by $\wtK^{n}$, gives
\begin{align} \label{kzero}
\wtK^n_{\bar\sigma^n_t}-\wtK^n_t=\int_{\rrb t,\bar\sigma^n_t\rrb}d\wtK^{n,r}_s + \int_{\llb t,\bar\sigma^n_t\llb} d\wtK^{n,g}_{s+} = 0.
\end{align}
If we take $\nu:=\tau \wedge \bar{\sigma}^n_t$ where $\tau\in\bSTFt$ is arbitrary, then
\begin{align*}
\wtY^n_t& = \EQ\big[\wtY^n_{\nu}-\int_{\rrb t, \nu\rrb} n(\wtY^n_s-R_s)^+\,d\wtGamma_s+\wt{K}^n_{\nu}-\wt{K}^n_t\,\big|\,\cF_t\big]\leq \EQ\big[\wtY^n_{\nu}|\,\cF_t\big] \nonumber\\
& = \EQ\big[\wtY^n_{\bar{\sigma}^n_t}\I_\seq{\tau > \bar{\sigma}^n_t}
          +\wtY^n_{\tau}\I_\seq{\tau \leq  \bar{\sigma}^n_t} \,|\,\cF_t\big] \nonumber\\
& \leq \EQ\big[P_{\bar{\sigma}^n_t}\I_\seq{\tau > \bar{\sigma}^n_t}+
     (\wtY^n\vee P\vee R)_{\tau}\I_\seq{\tau \leq \bar{\sigma}^n_t}\,|\,\cF_t\big] + \varepsilon\\
& \leq      \EQ\big[P_{\bar{\sigma}^n_t}\I_\seq{\tau > \bar{\sigma}^n_t}+
     (V^l\vee P\vee R)_{\tau}\I_\seq{\tau \leq \bar{\sigma}_t}\,|\,\cF_t\big] + \EQ\big[|\wtY^n_\tau-V^l_\tau|\,|\,\cF_t\big] + C\EQ\big[\I_{\rrb\bar\sigma_t, \bar\sigma^n_t \rrb}(\tau)|\,\cF_t\big] + \varepsilon
\end{align*}
where on the event $E_1:= \{\tau > \bar\sigma^n_t\}$ we have used the inequality $\wtY^n_{\bar{\sigma}^n_t} \leq P_{\bar{\sigma}^n_t} + \varepsilon$ while on the event $E_2:= \{\tau \leq \bar\sigma^n_t\}$ we have used the trivial inequality $\wtY^n_{\tau}\leq\wtY^n_{\tau}\vee P_{\tau}\vee R_{\tau}$.
By considering the limit superior in $n$ and using the conditional reverse Fatou lemma (see Theorem 2 in \cite{Z1998})  together with the upper-right-semicontinuity of $P$ and the monotone convergence theorem, we obtain
\begin{align} \label{eq222}
V^l_t \leq 	\EQ\big[P_{\bar{\sigma}_t}\I_\seq{\tau > \bar{\sigma}_t}+
     (V^l\vee P\vee R)_{\tau}\I_\seq{\tau \leq \bar{\sigma}_t}\,|\,\cF_t\big] +\varepsilon.
\end{align}
Our next goal is to show that $V^l$ can be omitted from the conditional expectation in \eqref{eq222}. For any $\tau \in \bSTFt$, equation \eqref{kzero} gives
\begin{align*}
\wtY^n_{\tau}=  \wtY^n_{\bar\sigma^n_\tau}-\int_{\rrb\tau,\bar\sigma^n_\tau\rrb} \wtZ^n_s\,d\whMM_s-\int_{\rrb \tau, \bar\sigma^n_\tau\rrb} n(\wtY^n_s-R_s)^+\,d\wtGamma_s.
\end{align*}

We now use similar arguments as in Step 2 in the proof of Theorem \ref{th6.1}. We observe that $\wtY^n_{\bar\sigma^n_\tau}\leq P_{\bar\sigma^n_\tau}+\varepsilon$ and, for all $(\omega,s, y) \in \Omega\times [0,T]\times \mathbb{R}$,
$$
(y-R_s)^+(\omega) \geq \I_{\llb 0, \bar\sigma_\tau\rrb}(s) (y-R_s)^+(\omega) \geq \I_{\llb 0, \bar\sigma_\tau\rrb}(s) (y-R_s)(\omega)
$$
where the function $g(\omega,s, y):=\I_{\llb 0, \bar\sigma_\tau\rrb}(s) (R_s-y)(\omega)$ is nonincreasing in $y$, for every $(\omega,s)\in \Omega\times [0,T]$.  By applying the comparison theorem for GBSDEs (see Proposition 3.1 in \cite{LLR2}) on the interval $\llb \tau, \bar\sigma^n_\tau\rrb$, we see that $\wtY^n \leq Y^n$ where $(Y^n,Z^n)$ solves the following linear BSDE
\begin{align*}
			 Y_\tau^n  & =  P_{\bar\sigma^n_\tau}  + \varepsilon-\int_{\rrb\tau,\bar\sigma^n_\tau\rrb} Z^n_s\,d\whMM_s-\int_{\rrb \tau, \bar\sigma^n_\tau\rrb} n(Y^n_s-R_s)\I_{\llb 0, \bar\sigma_\tau\rrb}(s) \,d\wtGamma_s.\\
			 & =  P_{\bar\sigma^n_\tau}  + \varepsilon-\int_{\rrb\tau,\bar\sigma^n_\tau\rrb} Z^n_s\,d\whMM_s+ \int_{\rrb \tau, \bar\sigma^n_\tau\rrb} n(R_s- Y^n_s)\,d\wtGamma^{\bar\sigma_\tau}_s.
\end{align*}
Since $\tau \leq \bar\sigma_\tau \leq \bar\sigma^n_\tau \leq T$, by solving the linear GBSDE we obtain
\begin{align*}
 Y_\tau^n& = \EQ\big[(P_{\bar\sigma^n_\tau}+ \varepsilon)\cE_{\tau,\bar\sigma_\tau}(-\wtGamma^n)+(\I_{\rrb \tau,\bar\sigma_\tau\rrb}R\cE_{\tau,\cdot}(-\wtGamma^n)\bcd \wtGamma^n)_{T}\, | \cF_\tau\big]\\
& \leq \EQ\big[P_{\bar\sigma_\tau}\cE_{\tau,\bar\sigma_\tau}(-\wtGamma^n)+(\I_{\rrb \tau,\bar\sigma_\tau\rrb}R\cE_{\tau,\cdot}(-\wtGamma^n)\bcd \wtGamma^n)_{T}\, | \cF_\tau\big]  + \EQ\big[[P_{\bar\sigma^n_\tau} - P_{\bar\sigma_\tau}]\cE_{\tau,\bar\sigma_\tau}(-\wtGamma^n)  |\cF_\tau\big] + \varepsilon
\end{align*}
where we have used the inequality $\cE_{\tau,\bar\sigma_\tau}(-\wtGamma^n) \leq 1$. The sequence $\cE_{\tau,\bar\sigma_\tau}(-\wtGamma^n)$ converges to $\I_{\{\tau=\bar\sigma_\tau\}}$ as $n\rightarrow \infty$ and, by the sub-additivity of the limit superior, the conditional reverse Fatou lemma and the dominated convergence theorem, we obtain
\begin{align*}
&\limsup_{\,n\rightarrow \infty}\, \EQ\big[[P_{\bar\sigma^n_\tau} - P_{\bar\sigma_\tau}]\cE_{\tau,\bar\sigma_\tau}(-\wtGamma^n) \,|\,\cF_\tau\big]\\
& \leq \limsup_{\,n\rightarrow \infty}\, \EQ\big[[P_{\bar\sigma^n_\tau} \cE_{\tau,\bar\sigma_\tau}(-\wtGamma^n)\,|\,\cF_\tau\big] - \lim_{\,n\rightarrow \infty}\, \EQ\big[P_{\bar\sigma_\tau}\cE_{\tau,\bar\sigma_\tau}(-\wtGamma^n)\,|\,\cF_\tau\big]\\
& \leq  \EQ\big[[\limsup_{\,n\rightarrow \infty} P_{\bar\sigma^n_\tau} - P_{\bar\sigma_\tau}]\I_{\{\tau=\bar\sigma^n_\tau\}}\,|\,\cF_\tau\big] \leq 0
\end{align*}
where the last inequality holds since $P$ is upper-right-semicontinuous along stopping times (see Remark B.3 in \cite{KQC2014}). For any fixed $\varepsilon > 0$, we deduce from the sub-additivity of the limit superior and Lemma \ref{lem3.4x} that
\begin{align*}
V^l_\tau\leq P_{\tau}\I_{\{\tau= \bar\sigma_\tau\}}+R_{\tau}\I_{\{\tau<\bar \sigma_\tau\}} + \varepsilon \leq (R\vee P)_\tau + \varepsilon
\end{align*}
and thus $V^l_\tau \leq (R\vee P)_\tau$ for every stopping time $\tau$ in $\bSTFt$, which gives the desired upper bound in \eqref{eq6.11b}.

\noindent {\it Step 3.} Since we always have that
\begin{align*}
\essinf_{\tau\in\bSTFt}\esssup_{\sigma\in\STFt}\EQ[X^l(\sigma,\tau)\,|\,\cF_t]\geq \esssup_{\sigma\in\STFt} \essinf_{\tau\in\bSTFt}\EQ[X^l(\sigma,\tau)\,|\,\cF_t]
\end{align*}
we obtain \eqref{eq6.7} by combining \eqref{eq6.11} with \eqref{eq6.12}.
\end{proof}

\noindent{\bf Acknowledgements.}  The research of R. Liu and M. Rutkowski was supported by the Australian Research Council Discovery Project DP200101550.


\end{document}